\pdfoutput=1%For ARXIV
\documentclass[a4paper,11pt]{article}
\usepackage{cite}
\usepackage[all]{xy}
\usepackage{amsmath,amsfonts,amssymb,amsthm,epsfig,amscd,comment,latexsym,psfrag}
\usepackage{CJK}
\usepackage{mathrsfs}
\usepackage{nicefrac,xspace,tikz}
\usepackage{arydshln}
\usepackage{stmaryrd}
\usepackage{graphicx,bm}
\usepackage{float}
\usepackage{amscd}
\usetikzlibrary{arrows}

\makeatletter

\newcommand{\Rmnum}[1]{\expandafter\@slowromancap\romannumeral #1@}
\makeatother
\usepackage{graphicx,amssymb,amsmath,bm,latexsym}
\allowdisplaybreaks

\begin{document}
\begin{titlepage}
\begin{center}
{\Large\bf Supersymmetric partition function hierarchies and character expansions}\vskip .2in
{\large Rui Wang$^{a,}$\footnote{wangrui@cumtb.edu.cn},
Fan Liu$^{b,}$\footnote{liufan-math@cnu.edu.cn},
Min-Li Li$^{b,}$\footnote{liml@cnu.edu.cn},
Wei-Zhong Zhao$^{b,}$\footnote{Corresponding author: zhaowz@cnu.edu.cn}} \vskip .2in
$^a${\em Department of Mathematics, China University of Mining and Technology,
Beijing 100083, China}\\
$^b${\em School of Mathematical Sciences, Capital Normal University,
Beijing 100048, China} \\
\begin{abstract}

We construct the supersymmetric $\beta$ and $(q,t)$-deformed Hurwitz-Kontsevich partition functions through
$W$-representations and present the corresponding character expansions with respect to the Jack and Macdonald
superpolynomials, respectively. Based on the constructed $\beta$ and $(q,t)$-deformed superoperators,
we further give the supersymmetric $\beta$ and $(q,t)$-deformed partition function hierarchies
through $W$-representations. We also present the generalized super Virasoro constraints,
where the constraint operators obey the generalized super Virasoro algebra and null super 3-algebra.
Moreover, the superintegrability for these (non-deformed) supersymmetric hierarchies
is shown by their character expansions, i.e., $<character>\sim character$.

\end{abstract}
\end{center}

{\small Keywords: Matrix Models, Supersymmetry}

\end{titlepage}

\section{Introduction}

$W$-representation of matrix model realizes the partition function by acting on elementary functions
with exponents of the given $W$-operator \cite{Shakirov2009}. A considerable amount is already known
about $W$-representations for matrix models. Very recently it was shown that
the spectral curve can be extracted from the $W$-representation of matrix model \cite{2210.09993}.
The spectral curve was associated with a peculiar part $\hat W^{spec}$ of the $\hat W$-operator.
As the generalizations of matrix models from matrices to tensor,
tensor models provide the analytical tool for the study of random geometries in three and more dimensions.
Hence they are serious candidates for a theory of quantum gravity. The studies of $W$-representations
for tensor models have been carried out. It was found that there are the $W$-representations for
the Gaussian tensor model \cite{Itoyama2020} and (fermionic) rainbow tensor models \cite{LY,Kang2021}.
Due to the $W$-representations, it allows the correlators of these models to be exactly calculated.
In addition, there have also been attempts to investigate $W$-representations of supereigenvalue models.
For the supereigenvalue model in the Ramond sector, its free energy depends on Grassmann couplings only up to quadratic order \cite{Osuga}.
The $W$-representation for this supereigenvalue model was presented in Ref.\cite{Chen2020}.
The supereigenvalue model in the Neveu-Schwarz sector describes the coupling between
$(2, 4m)$ superconformal models and world-sheet supergravity \cite{Zadra}.
For the Gaussian and chiral supereigenvalue models in the Neveu-Schwarz sector, it was noted that
there are the so called generalized $W$-representations. More precisely, they can be expressed as the infinite sums of the homogeneous operators
acting on the elementary functions \cite{Wang2020}. An important feature of these models is that the compact expressions of the correlators
can be derived from such generalized $W$-representations.

The superintegrability for matrix models has attracted much attention (see \cite{Mironovsummary} and the references therein, \cite{2203.03869}-\cite{Al}).
Here the superintegrability means that for the character expansions of the matrix models, the average of a properly chosen
symmetric function is proportional to ratios of symmetric functions on a proper locus, i.e., $< character > \sim character$.
$W$-representations have contributed to our understanding of the superintegrability.
For a wide range of superintegrable matrix models, their character expansions can be derived from the corresponding $W$-representations.
What is also worth noticing is the Virasoro constraints for matrix models.
They were used to analyze the character expansions of matrix models as well \cite{Rashkov, LY}.
The Hurwitz-Kontsevich (HK) matrix model with $W$-representation is an important superintegrable matrix model which can be used to
describe the Hurwitz numbers and Hodge integrals over the moduli space of complex curves \cite{Shakirov2009,Goulden97,KHurwitz}.
Its character expansions with respect to the Schur functions can be easily derived from the $W$-representation.
The HK partition function was extended to the $\beta$ and $(q,t)$-deformed cases \cite{Wang2022,qtHurwitz}.
The superintegrability for these deformed HK models was confirmed by the character expansions with respect to the Jack and Macdonald
polynomials, respectively. Moreover, the superintegrable $\beta$ and $(q,t)$-deformed partition function hierarchies have been constructed,
where the $\beta$ and $(q,t)$-deformed Hurwitz operators play a fundamental role in the $W$-representations.

Quite recently, it was shown that the partition function hierarchies constructed in Ref. \cite{Wang2022} can be described by the two-matrix
model that depends on two (infinite) sets of variables and an external matrix \cite{Al,2301.04107,2301.11877}. Their generalizations were
realized by $W$-representations associated with infinite commutative families of generators of $w_{\infty}$ algebra.
They are the special cases of skew hypergeometric Hurwitz $\tau$-functions which are $\tau$-functions of the Toda lattice hierarchy of the
skew hypergeometric type. The multi-matrix representation as well as their $\beta$-deformations were provided in Ref.\cite{2301.11877}.
$W$-representations for ($\beta$-deformed) multi-character partition functions were constructed in Ref. \cite{2301.12763}.
They involve a generic number of sets of time variables and the integral representations for such kind of partition functions are given
by tensor models and multi-matrix models with multi-trace couplings in some special cases.

The goal of this paper is to make a step towards the supersymmetric case, i.e., the superintegrability for the partition functions depending
on bosonic and fermionic (or Grassmann) variables. We shall construct the supersymmetric partition function hierarchies through $W$-representations
and analyze the superintegrability.

This paper is organized as follows.
In section 2, the $\beta$-deformed Hurwitz operator is extended to the supersymmetric case. We construct the supersymmetric $\beta$-deformed
HK partition functions and present their character expansions with respect to the Jack superpolynomials.
In sections 3, based on the $\beta$-deformed Hurwitz superoperators, we give the supersymmetric $\beta$-deformed hierarchy for
the partition functions through $W$-representations.  The superintegrability is shown by the character expansions with respect to the Jack superpolynomials.
We also construct the constraints for the supersymmetric $\beta$-deformed partition functions.
The remarkable feature of constraint operators is that they yield the new infinite-dimensional super algebra and null super 3-algebra.
In section 4, we construct the $(q,t)$-deformed Hurwitz superoperators and present the supersymmetric $(q,t)$-deformed HK partition functions.
Then the supersymmetric $(q,t)$-deformed partition function hierarchy is constructed through $W$-representations and the desired constraints
are presented as well. Moreover, the superintegrability is shown by the character expansions with respect to the Macdonald superpolynomials.
We end this paper with the conclusions in section 5.

%%%%%%%%%%%%%%%%%%%%%%%%%%%%%%%%%%%%%%%%%%%%%%%%%%%%%%%%%%%%%%%%%%%%%%%%%%%%%%%%%%%%%%%%%%%%%
\section{Supersymmetric $\beta$-deformed HK partition functions}
%%%%%%%%%%%%%%%%%%%%%%%%%%%%%%%%%%%%%%%%%%%%%%%%%%%%%%%%%%%%%%%%%%%%%%%%%%%%%%%%%%%%%%%%%%%%%
A superpartition $\Lambda\mapsto (n|m)$ of bosonic degree
$n=|\Lambda|=\sum_{i=1}^{l(\Lambda)}\Lambda_i$ and fermionic degree $m$  is a pair of partitions written as \cite{Desrosiers2001}-\cite{Desrosiers2006}
$$\Lambda=(\Lambda^a;\Lambda^s)=(\Lambda_1,\Lambda_2,\cdots,\Lambda_m;
\Lambda_{m+1},\Lambda_{m+2},\cdots,\Lambda_{l(\Lambda)}),$$
where $\Lambda_i$, $i=1,\cdots,l(\Lambda)$ are integers,
$\Lambda_1>\Lambda_2>\cdots>\Lambda_m\geq 0$ and  $\Lambda_{m+1}\geq\Lambda_{m+2}\geq\cdots\geq\Lambda_{l(\Lambda)}>0$.

Note that the superpartitions of degree $(n|0)$ are regular partitions. In the following, we denote
$\Lambda^*=(\Lambda^a;\Lambda^s)^+$ as the  partition obtained by reordering the concatenation of
entries of $\Lambda^a$ and $\Lambda^s$ in non-increasing order, $\Lambda^\circledast=(\Lambda^a+1^m;\Lambda^s)^+$,
where $\Lambda^a+1^m$ is the partition obtained by adding one to each entry of
$\Lambda^a$.

The diagram of $\Lambda$ is given by drawing the diagram of
$\Lambda^\circledast$ and then replacing the cells of $\Lambda^\circledast/\Lambda^*$ by circles.
To make it a little bit more clear, let us now take $\Lambda=(3,2,0;6,4,2)\mapsto (17|3)$ as an example,
then $\Lambda^*=(6,4,3,2,2)$, $\Lambda^\circledast=(6,4,4,3,2,1)$:
\begin{figure}[H]
  \centering
  \includegraphics[width=7.5cm]{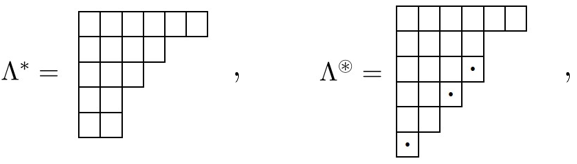}
\end{figure}
{\noindent where the box with a dot is in the skew diagram $\Lambda^\circledast/\Lambda^*$.}
The diagram of $\Lambda$ is represented as
\begin{figure}[H]
  \centering
  \includegraphics[width=7.5cm]{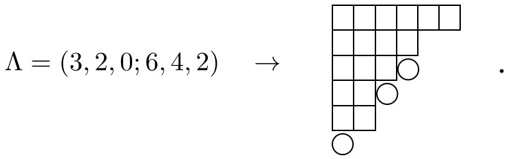}
\end{figure}

For two superpartitions $\Lambda$ and $\Omega$, we say
$\Lambda\subseteq \Omega$ if $\Lambda^*\subseteq \Omega^*$ and $\Lambda^\circledast\subseteq \Omega^\circledast$.

It is known that the Jack superpolynomials $J_{\Lambda}$ are eigenfunctions of two commuting
superoperators \cite{Desrosiers2003,Desrosiers2012}
\begin{eqnarray}
D&=&\frac{1}{2}\sum_{i=1}^{N}x_i^2\frac{\partial^2}{\partial x_i^2}
+\beta \sum_{1\leq i\neq j\leq N}\frac{x_ix_j}{x_i-x_j}\left(\frac{\partial}{\partial x_i}
-\frac{\theta_i-\theta_j}{x_i-x_j}\frac{\partial}{\partial \theta_i}\right),\nonumber\\
\Delta&=&\sum_{i=1}^{N}x_i\theta_i\frac{\partial}{\partial x_i}\frac{\partial}{\partial \theta_i}
+\beta \sum_{1\leq i\neq j\leq N}\frac{x_i\theta_j+x_j\theta_i}{x_i-x_j}\frac{\partial}{\partial \theta_i},
\end{eqnarray}
and
\begin{eqnarray}
DJ_{\Lambda}&=&\left(\sum_{j=1}^{l({\Lambda'}^*)}(j-1){\Lambda_j'}^*-\beta\sum_{i=1}^{
l({\Lambda}^*)}(i-1)\Lambda_i^*\right)J_{\Lambda},\nonumber\\
\Delta J_{\Lambda}&=&(|\Lambda^a|-\beta|{\Lambda'}^a|)J_{\Lambda},
\end{eqnarray}
where $x_i$ and $\theta_i$, $i=1,\cdots, N$, are the bosonic and fermionic variables, respectively,
$\Lambda'$ is the conjugate of $\Lambda$ whose diagram is given by reflecting the diagram of $\Lambda$
with respect to the main diagonal.

Let us introduce the superoperators
\begin{equation}\label{wcircle}
\mathcal{W}^{\circ}=D+\beta NE+\delta(\Delta+\beta N\bar E),\ \
\end{equation}
where we take $\circ\in \{*, \circledast\}$ and
$\delta=\left\{
\begin{array}{ll}
 1, & \hbox{$\circ=\circledast$;} \\
0, & \hbox{$\circ=*$,}
\end{array}
\right.$ in this paper,
 $E=\sum_{i=1}^{N}x_i\frac{\partial}{\partial x_i}$ and $\bar E=\sum_{i=1}^{N}\theta_i\frac{\partial}{\partial \theta_i}$.

Then we have
\begin{eqnarray}\label{dW0action}
\mathcal{W}^{\circ} J_{\Lambda}=\sum_{(i,j)\in \Lambda^{\circ}}c(N;i,j)J_{\Lambda},
\end{eqnarray}
where $c(N;i,j)=j-1+\beta (N-i+1)$.

The superoperators ${\mathcal{W}}^*$ and ${\mathcal{W}}^\circledast$ are actually commutative,
since they are expressed by the mutually commuting conserved charges in the supersymmetric trigonometric
Calogero-Moser-Sutherland (CMS) model.

Let $p_k=\sum_{i}x_i^k$, $k=1,2,\cdots$, and  $\tilde p_l=\sum_{i}\theta_ix_i^l$,
$l=0,1,2,\cdots$, be the bosonic and fermionic power sums, respectively.
In terms of the variables $\{p\}=\{p_1,p_2,\cdots\}$ and $\{\tilde p\}=\{\tilde p_0,\tilde p_1,\tilde p_2,\cdots\}$,
the superoperators (\ref{wcircle}) can be written as
\begin{eqnarray}\label{shkope}
\mathcal{W}^\circ&=&\frac{1}{2}\sum_{k,l=1}^{\infty}(\beta(k+l)p_{k}p_l
\frac{\partial}{\partial p_{k+l}}+klp_{k+l}\frac{\partial}{\partial p_k}\frac{\partial}{\partial p_l})
+\sum_{k,l=1}^{\infty}(\beta lp_{k}\tilde p_l
\frac{\partial}{\partial \tilde p_{k+l}}\nonumber\\
&&+kl \tilde p_{k+l}\frac{\partial}{\partial \tilde p_k}
\frac{\partial}{\partial p_l})+\frac{1}{2}\sum_{k=1}^{\infty}\big((1-\beta)(k-1)
+2\beta N\big)k\big(p_{k}\frac{\partial}{\partial p_k}
+\tilde p_{k}\frac{\partial}{\partial \tilde p_k}\big)\nonumber\\
&&+\delta\big(\sum_{n\geq 0}((1-\beta)n+\beta N)\tilde p_n
\frac{\partial}{\partial \tilde p_n}+\sum_{m\geq 0,n>0}(n\tilde p_{m+n}
\frac{\partial}{\partial \tilde p_m}\frac{\partial}{\partial  p_n}+\beta p_n
\tilde p_m\frac{\partial}{\partial \tilde p_{m+n}})\big).
\end{eqnarray}

Setting to zero all the fermionic variables $\{\tilde p\}$ in (\ref{shkope}), both the
superoperators ${\mathcal{W}}^*$ and ${\mathcal{W}}^\circledast$ reduce to the
$\beta$-deformed Hurwitz operator \cite{Wang2022}
\begin{eqnarray}\label{Hoperator}
\hat W_0&=&\frac{1}{2}\sum_{k,l=1}^{\infty}\big(\beta(k+l)p_{k}p_l
\frac{\partial}{\partial p_{k+l}}+klp_{k+l}\frac{\partial}{\partial p_k}\frac{\partial}{\partial p_l}\big)\nonumber\\
&&+\frac{1}{2}\sum_{k=1}^{\infty}\big((1-\beta)(k-1)+2\beta N\big)kp_{k}\frac{\partial}{\partial p_k}.
\end{eqnarray}
Thus we call $\mathcal W^\circ$ the  $\beta$-deformed Hurwitz superoperators.

We give the supersymmetric $\beta$-deformed HK partition functions through $W$-representations
\begin{equation}\label{SHKM}
{\mathcal Z}^{\circ}\{p,\tilde p\}
=e^{\tau \mathcal{W}^{\circ}} e^{\beta e^{-\tau N}(p_1+\tilde p_1\theta)},
\end{equation}
where $\tau$ and $\theta$ are the bosonic and fermionic parameters, respectively.

The Cauchy formula for the Jack superpolynomials is \cite{Desrosiers2007}
\begin{equation}\label{cauchy}
e^{\beta\sum_{k=1}^{\infty}(\frac{p_kg_k}{k}+\tilde p_{k-1} \tilde g_{k-1}) }=\sum_{\Lambda}
\frac{1}
{\langle J_{\Lambda}, J_{\Lambda}\rangle_{\beta}}J_{\Lambda}\{p,\tilde p\}J_{\Lambda}\{g,\tilde g\},
\end{equation}
where $\langle J_{\Lambda}, J_{\Lambda}\rangle_{\beta}=(-1)^{\frac{m(m-1)}{2}}\beta^{-m}\prod_{s\in \mathcal{B}\Lambda}
\frac{h^{up}_{\Lambda}(s;\beta)}{h^{lo}_{\Lambda}(s;\beta)}$, $m$ is the fermionic degree of $\Lambda$, $\mathcal{B}\Lambda$
is the set of squares in the diagram of $\Lambda$ that do not belong at the same time to a fermionic row and
a fermionic column,
$h^{up}_{\Lambda}(s;\beta)=\beta^{-1}(a_{\Lambda^*}(s)+1)+l_{\Lambda^{\circledast}}(s)$,
$h^{lo}_{\Lambda}(s;\beta)=\beta^{-1}a_{\Lambda^{\circledast}}(s)
+l_{\Lambda^*}(s)+1$, for $s=(i,j)$ in the partition $\lambda$, the arm length $a_{\lambda}(s)=\lambda_i-j$
and leg length $l_{\lambda}(s)=\lambda^{'}_j-i$.

Taking $g_k=e^{-\tau N}\delta_{k,1}$ and $\tilde g_k=e^{-\tau N}\theta \delta_{k,1}$ in (\ref{cauchy})
and using (\ref{dW0action}), we may write the character expansions of (\ref{SHKM}) as
\begin{eqnarray}
\mathcal Z^\circ\{p,\tilde p\}
=\sum_{\Lambda}e^{\tau\sum_{(i,j)\in \Lambda^\circ}c(N;i,j)}
\frac{1}
{\langle J_{\Lambda}, J_{\Lambda}\rangle_{\beta}}J_{\Lambda}\{p,\tilde p\}
J_{\Lambda}\{g_k=e^{-\tau N}\delta_{k,1},\tilde g_k=e^{-\tau N}\theta\delta_{k,1}\}.
\end{eqnarray}

The non-deformed partition functions follow from (\ref{SHKM}) by taking $\beta=1$,
\begin{eqnarray}\label{SHKPF}
Z^\circ\{p,\tilde p\}
&=&e^{\tau{W}^\circ}e^{ e^{-\tau N}(p_1+\tilde p_1\theta)}\nonumber\\
&=&\sum_{\Lambda}e^{\tau c_{\Lambda^\circ}}
\frac{1}
{\langle S_{\Lambda}^{Jack}, S_{\Lambda}^{Jack}\rangle}S^{Jack}_{\Lambda}\{p,\tilde p\}
S_{\Lambda}^{Jack}\{g_k=e^{-\tau N}\delta_{k,1},\tilde g_k=e^{-\tau N}\theta\delta_{k,1}\},
\end{eqnarray}
where
\begin{eqnarray}\label{sshkope}
{W}^\circ&=&\frac{1}{2}\sum_{k,l=1}^{\infty}\big((k+l)p_{k}p_l
\frac{\partial}{\partial p_{k+l}}+klp_{k+l}\frac{\partial}{\partial p_k}\frac{\partial}{\partial p_l}\big)\nonumber\\
&&+\sum_{k,l=1}^{\infty}\big( lp_{k}\tilde p_l
\frac{\partial}{\partial \tilde p_{k+l}}+kl \tilde p_{k+l}\frac{\partial}{\partial \tilde p_k}\frac{\partial}{\partial p_l}\big)
+N\sum_{k=1}^{\infty}\big(kp_{k}\frac{\partial}{\partial p_k}+k\tilde p_{k}\frac{\partial}{\partial \tilde p_k}\big)\nonumber\\
&&+\delta\big(N\sum_{n\geq 0}\tilde p_n
\frac{\partial}{\partial \tilde p_n}+\sum_{m\geq 0,n>0}(n\tilde p_{m+n}
\frac{\partial}{\partial \tilde p_m}\frac{\partial}{\partial  p_n}+ p_n
\tilde p_m\frac{\partial}{\partial \tilde p_{m+n}})\big),
\end{eqnarray}
$c_{\lambda}=\sum_{(i,j)\in \lambda}(N-i+j)$, $S_{\Lambda}^{Jack}$ is the Schur-Jack superpolynomial \cite{Blondeau} and
$\langle S_{\Lambda}^{Jack}, S_{\Lambda}^{Jack}\rangle=(-1)^{\frac{m(m-1)}{2}}\prod_{s\in \mathcal{B}\Lambda}
\frac{h^{up}_{\Lambda}(s;1)}{h^{lo}_{\Lambda}(s;1)}$ with $m$ the fermionic degree of $\Lambda$.

Setting to zero all the fermionic variables in (\ref{SHKPF}), we recover
the HK model with $W$-representation \cite{Shakirov2009,Goulden97,KHurwitz}
\begin{eqnarray}\label{HKPF}
Z_{0}\{p\}=e^{\tau {\tilde W}_0}\cdot e^{p_1/e^{\tau N}}
=\sum_{\lambda}e^{\tau c_{\lambda}}S_\lambda\{p_k=e^{-\tau N}\delta_{k,1}\} S_{\lambda}\{p\},
\end{eqnarray}
where the Hurwitz operator ${\tilde W}_0$ is given by (\ref{Hoperator}) with $\beta=1$, $S_{\lambda}$ is the Schur function.

The matrix model representation of (\ref{HKPF}) is given by \cite{Shakirov2009}
\begin{eqnarray}
Z_{0}\{p\}=\int_{N\times N} \sqrt{{\rm det}\left(\frac{{\rm sinh}(\frac{\phi\otimes I-I\otimes\phi}{2})}
{\frac{\phi\otimes I-I\otimes\phi}{2}} \right)}d\phi e^{-\frac{1}{2t}{\rm Tr}\phi^2
-\frac{N}{2}{\rm Tr}\phi-\frac{1}{6}tN^3+\frac{1}{24}tN+{\rm Tr}(e^{\phi}\psi)},
\end{eqnarray}
where  $\psi$ is an $N\times N$ matrix and the time variables $p_k={\rm Tr}\psi^k$.

%%%%%%%%%%%%%%%%%%%%%%%%%%%%%%%%%%%%%%%%%%%%%%%%%%%%%%%%%%%%%%%%%%%%%%%%%%%%%%%%%%%%%%%%%%%%%
\section{Supersymmetric $\beta$-deformed partition function hierarchy}
%%%%%%%%%%%%%%%%%%%%%%%%%%%%%%%%%%%%%%%%%%%%%%%%%%%%%%%%%%%%%%%%%%%%%%%%%%%%%%%%%%%%%%%%%%%%%
%%%%%%%%%%%%%%%%%%%%%%%%%%%%%%%%%%%%%%%%%%%%%%%%%%%%%%%%%%%%%%%%%%%%%%%%%%%%%%%%%%%%%%%%%%%%%
\subsection{The negative branch of hierarchy}
%%%%%%%%%%%%%%%%%%%%%%%%%%%%%%%%%%%%%%%%%%%%%%%%%%%%%%%%%%%%%%%%%%%%%%%%%%%%%%%%%%%%%%%%%%%%%

Let us define the bosonic operators
\begin{eqnarray}
{\mathcal W}^{(n_1,n_2)}_{-1}(\vec{u})&=&\prod_{i=n_1+1}^{n_1+n_2}{\rm ad}_{{\mathcal{W}}^*(u_i)}
\prod_{j=1}^{n_1}{\rm ad}_{{\mathcal{W}}^\circledast(u_j)}p_1\nonumber\\
&=&[{\mathcal{W}}^*(u_{n_1+n_2}),
\cdots,[{\mathcal{W}}^*(u_{n_1+1}),[{\mathcal{W}}^\circledast(u_{n_1}),
\cdots,[{\mathcal{W}}^\circledast(u_{1}),p_1]\cdots]],
\end{eqnarray}
where $(n_1,n_2)\in \mathbb{N}^2\setminus (0,0)$, $\vec{u}=(u_1,u_2,\cdots,u_{n_1+n_2})$,
the notation ${\rm ad}_A B=[A, B]=AB-BA$ and $\mathcal W^{\circ}(u_i)$ are given by replacing $N$
in the $\beta$-deformed Hurwitz superoperators (\ref{shkope}) with an arbitrary parameter $u_i$.

The actions of ${\mathcal W}^{(n_1,n_2)}_{-1}(\vec{u})$
on the Jack superpolynomials are
\begin{eqnarray}\label{ew-1naction}
\mathcal{W}_{-1}^{(n_1,n_2)}(\vec{u})J_{\Lambda}
=\sum_{\Omega}C(n_1,n_2;\Omega,\Lambda)
\frac{\langle p_1J_{\Lambda},J_{\Omega}\rangle_{\beta}}
{\langle J_{\Omega},J_{\Omega}\rangle_{\beta}} J_{\Omega},
\end{eqnarray}
where $C(n_1,n_2;\Omega,\Lambda)=\prod_{(i_1,j_1)\in\Omega^\circledast/\Lambda^\circledast}
\prod_{r_1=1}^{n_1}c(u_{r_1};i_1,j_1)
\prod_{(i_2,j_2)\in\Omega^*/\Lambda^*}
\prod_{r_2=n_1+1}^{n_1+n_2}c(u_{r_2};i_2,j_2)$, the sum is over the superpartitions $\Omega$ satisfying
$\Lambda\subseteq \Omega$ and
$|\Omega^*/\Lambda^*|=|\Omega^\circledast/\Lambda^\circledast|=1$.

The bosonic operators $\mathcal{W}_{-k}^{(n_1,n_2)}(\vec{u})$
$(k\geq 2)$ are defined as
\begin{eqnarray}\label{ew-knku}
{\mathcal{W}}_{-k}^{(n_1,n_2)}(\vec{u})
=\frac{1}{(k-1)!}{\rm ad}_{{\mathcal{W}}_{-1}^{(n_1,n_2+1)}(\vec{u})}^{k-1}
{\mathcal{W}}_{-1}^{(n_1,n_2)}
(\vec{u}),\ \ k\geq 2.
\end{eqnarray}
There are the actions
\begin{eqnarray}\label{ew-knaction}
\mathcal{W}_{-k}^{(n_1,n_2)}(\vec{u})J_{\Lambda}
=\sum_{\Omega}C(n_1,n_2;\Omega,\Lambda)\frac{\langle p_kJ_{\Lambda},J_{\Omega}\rangle_{\beta}}
{\langle J_{\Omega},J_{\Omega}\rangle_{\beta}} J_{\Omega},
\end{eqnarray}
where the sum is over the superpartitions $\Omega$ satisfying
$\Lambda\subseteq \Omega$ and
$|\Omega^*/\Lambda^*|=|\Omega^\circledast/\Lambda^\circledast|=k$.

We may introduce the fermionic operators $\mathcal{V}_{-k}^{(n_1,n_2)}(\vec{u})$,
$(n_1,n_2)\in \mathbb{N}^2\setminus (0,0)$, as follows:
\begin{equation}\label{ev-kn}
\mathcal{V}_{-k}^{(n_1,n_2)}(\vec{u})=
\left\{
  \begin{array}{ll}
\prod_{i=n_1+1}^{n_1+n_2}{\rm ad}_{{\mathcal{W}}^*(u_i)}
\prod_{j=1}^{n_1}{\rm ad}_{{\mathcal{ W}}^\circledast(u_j)}\tilde p_k,
& \hbox{$k=0,1$;} \\
\\
\frac{1}{k-1}[\mathcal{V}_{-1}^{(n_1,n_2+1)}(\vec{u}),
\mathcal{W}_{-k+1}^{(n_1,n_2)}(\vec{u})], & \hbox{$k\geq 2$.}
  \end{array}
\right.
\end{equation}
By means of the action (\ref{ew-knaction}) and relation $\frac{1}{k-1}[\mathcal{V}_{-1}^{(0,1)}(\vec{u}),
p_{k-1}]=\tilde p_k$,
it is not difficult to obtain that
\begin{eqnarray}\label{ev-knaction}
\mathcal{V}_{-k}^{(n_1,n_2)}(\vec{u})J_{\Lambda}
=\sum_{\Omega}C(n_1,n_2;\Omega,\Lambda)\frac{\langle \tilde p_kJ_{\Lambda},J_{\Omega}\rangle_{\beta}}
{\langle J_{\Omega},J_{\Omega}\rangle_{\beta}} J_{\Omega},
\end{eqnarray}
where the sum is over the superpartitions $\Omega$ satisfying
$\Lambda\subseteq \Omega$ and
$|\Omega^*/\Lambda^*|=k$, $|\Omega^\circledast/\Lambda^\circledast|=k+1$.

By means of the bosonic operators (\ref{ew-knku}) and fermionic operators (\ref{ev-kn}),
we give the negative branch of supersymmetric $\beta$-deformed hierarchy
\begin{eqnarray}\label{eznm}
\mathcal{Z}_{-n_1,-n_2}\{\vec u;p,\tilde p|g,\tilde g\}=
e^{\beta\sum_{k=1}^{\infty}({\mathcal{W}}_{-k}^{(n_1,n_2)}(\vec{u})
\frac{g_k}{k}+{\mathcal{V}}_{-k+1}^{(n_1,n_2)}(\vec{u})\tilde g_{k-1})}\cdot 1.
\end{eqnarray}
Since the operators ${\mathcal{W}}_{-k}^{(n_1,n_2)}(\vec{u})$ and ${\mathcal{V}}_{-k+1}^{(n_1,n_2)}(\vec{u})$ in (\ref{eznm})
are defined by the nested commutators, we call a family of partition functions (\ref{eznm}) ``hierarchy" here.

In order to derive the character expansion of (\ref{eznm}), let us recall the evaluation formulas for
the Jack superpolynomials. We denote $\Lambda=(\Lambda^a;\Lambda^s)$ as the superpartition
of fermionic degree $m$.
The evaluation $E_u$ on the power sum basis $p_{\Lambda}:=\prod_{i=1}^m \tilde p_{\Lambda_i^a}\prod_{j=m+1}^{l(\Lambda)} p_{\Lambda_j^s}$
is defined as \cite{Desrosiers2012}
\begin{equation}
E_u(p_\Lambda)=u^{l(\Lambda^s)}S_{\Lambda^a-\delta_m}(1,\cdots, 1),
\end{equation}
where $\delta_m=(m-1,m-2,\cdots,1)$, $S_{\Lambda^a-\delta_m}$ is the Schur polynomial associated with the partition $\Lambda^a-\delta_m$.

Then the evaluation formulas on the Jack superpolynomial $J_{\Lambda}$ are given by \cite{Desrosiers2012}
\begin{eqnarray}
E_{u}(J_{\Lambda})&=&E'_\beta(J_{\Lambda})
\prod_{(i,j)\in \Lambda^\circledast/\delta_{m+1}}c(u;i,j),\ \ m\geq 0,\label{ev2}\nonumber\\
\tilde E_{u}(J_{\Lambda})&:=&E_u(((-1)^{m-1}
\partial_{\theta_{N}}J_{\Lambda})|_{x_{N=0}})\nonumber\\
&=&E'_\beta(J_{\Lambda})
\prod_{(i,j)\in \Lambda^*/\delta_m}c(u;i,j), \ \ m\geq 1,\label{ev1}
\end{eqnarray}
where
\begin{equation}
E'_{\beta}(J_{\Lambda})=\left.\left(\Delta_m^{-1}\frac{\partial}{\partial \theta_m}\cdots \frac{\partial}{\partial \theta_1}
J_{\Lambda}\right)\right|_{\substack{x_1=\cdots=x_m=0\\x_{m+1}^k+\cdots+x_N^k=\beta^{-1}\delta_{k,1}}}
=\beta^{-|\Lambda^*/\delta_m|}\prod_{s\in \mathcal{B}\Lambda}h^{lo}_{\Lambda}(s;\beta)^{-1},
\end{equation}
$\Delta_m=\prod_{1\leq i<j\leq N}(x_i-x_j)$.

Note that when $m=0$ in (\ref{ev2}), it gives the evaluation formula for the Jack polynomial $J_{\lambda}$ associated
with the regular partition $\lambda$ \cite{Macdonaldbook}
\begin{eqnarray}\label{hook}
J_{\lambda}\{p_k=u\}
=J_{\lambda}\{p_k=\beta^{-1}\delta_{k,1}\}
\prod_{(i,j)\in \lambda}c(u;i,j).
\end{eqnarray}

Using the actions (\ref{ew-knaction}), (\ref{ev-knaction}) and evaluation formulas (\ref{ev1}) and (\ref{hook}),
we obtain the character expansion of (\ref{eznm})
\begin{eqnarray}\label{ceznm}
\mathcal{Z}_{-n_1,-n_2}\{\vec u;p,\tilde p|g,\tilde g\}
&=&\sum_{\lambda}\prod_{r=1}^{n_1+n_2}\frac{E_{u_r}(J_{\lambda})}{E'_{\beta}(J_{\lambda})}
\frac{J_{\lambda}\{p\}J_{\lambda}\{g\}}{\langle J_{\lambda},J_{\lambda}\rangle_{\beta}}
+\sum_{\Lambda}\frac{1}{E'_\beta(J_{\Lambda})^{n_1+n_2}}\prod_{r_1=1}^{n_1}\prod_{r_2=n_1+1}^{n_1+n_2}
\nonumber\\
&&\cdot E_{u_{r_1}}(J_{\Lambda})\frac{E_{u_{r_1}}(J_{\delta_{m+1}})}
{E'_\beta(J_{\delta_{m+1}})}
\tilde E_{u_{r_2}}(J_{\Lambda})\frac{E_{u_{r_2}}(J_{\delta_{m}})}
{E'_{\beta}(J_{\delta_{m}})}J_{\Lambda}\{p,\tilde p\}\frac{J_{\Lambda}\{g,\tilde g\}}{\langle J_{\Lambda}, J_{\Lambda}\rangle_{\beta}},
\end{eqnarray}
where $\lambda$ is the regular partition, $\Lambda$ is the superpartition of fermionic degree $m>0$.

The supersymmetric $\beta$-deformed  partition functions (\ref{eznm}) satisfy the constraints
\begin{equation}\label{gconb}
{\mathbb W}_k\mathcal{Z}_{-n_1,-n_2}\{\vec u;p,\tilde p|g,\tilde g\}=0,
\end{equation}
and
\begin{equation}\label{gconf}
{\mathbb V}_l\mathcal{Z}_{-n_1,-n_2}\{\vec u;p,\tilde p|g,\tilde g\}=0,
\end{equation}
where the constraint operators are given by
\begin{eqnarray}
&&{\mathbb W}_k=\beta^{-1}k\frac{\partial}{\partial g_k}-{\mathcal{W}}_{-k}^{(n_1,n_2)}(\vec{u}),  \ \ k\geq 1,\nonumber\\
&&{\mathbb V}_l=\beta^{-1}\frac{\partial}{\partial \tilde g_l}-{\mathcal{V}}_{-l}^{(n_1,n_2)}(\vec{u}), \ \ l\geq 0,
\end{eqnarray}
they yield null super algebra.

Multiplying by $g_k$ and then taking the sum over $k$ in (\ref{gconb}), we have
\begin{equation}\label{l0wk}
(\beta^{-1}l_0\{g\}-\sum_{k=1}^{\infty}g_k{\mathcal{W}}_{-k}^{(n_1,n_2)}(\vec{u}))
\mathcal{Z}_{-n_1,-n_2}\{\vec u;p,\tilde p|g,\tilde g\}=0,
\end{equation}
where $l_0\{g\}=\sum_{k=1}^{\infty}kg_k\frac{\partial}{\partial g_k}$.

Similarly, from (\ref{gconf}) we have
\begin{equation}\label{l0vk}
(\beta^{-1}\tilde l_0\{\tilde g\}-\sum_{k=1}^{\infty}k\tilde g_k{\mathcal{V}}_{-k}^{(n_1,n_2)}(\vec{u}))
\mathcal{Z}_{-n_1,-n_2}\{\vec u;p,\tilde p|g,\tilde g\}=0,
\end{equation}
where $\tilde l_0\{\tilde g\}=\sum_{k=1}^{\infty}k\tilde g_k\frac{\partial}{\partial \tilde g_k}$.

Let us pause here to recall the Gaussian hermitian one-matrix model \cite{Shakirov2009,AMironov1705}
\begin{eqnarray}\label{GPF}
Z_G\{p\}&=&2^{-\frac{N}{2}}\pi^{-\frac{N^2}{2}}\int_{N\times N} dM e^{-\frac{1}{2}{\rm Tr}M^{2}+\sum_{k=1}^{\infty}\frac{p_k}{k}{\rm Tr}M^k}\nonumber\\
&=&e^{{\hat W}_{-2}/2}\cdot 1=\sum_{\lambda} S_{\lambda}\{p_k=\delta_{k,2}\}\frac{S_{\lambda}\{p_k=N\}}{S_{\lambda}\{p_k=\delta_{k,1}\}}S_{\lambda}\{p\},
\end{eqnarray}
where
\begin{eqnarray}
{\hat W}_{-2}&=&\sum_{k,l=1}^{\infty}\big((k+l-2)p_{k}p_l
\frac{\partial}{\partial p_{k+l-2}}+klp_{k+l+2}\frac{\partial}{\partial p_k}\frac{\partial}{\partial p_l}\big)\nonumber\\
&&+2N\sum_{k=1}^{\infty}kp_{k+2}
\frac{\partial}{\partial p_{k}}+N(p_1^2+Np_2),
\end{eqnarray}
$\lambda$ is the regular partition.

There are the well known Virasoro constraints for (\ref{GPF})
\begin{equation}\label{VCGHM}
L_nZ_G\{p\}=0,
\end{equation}
where the Virasoro constraint operators are
\begin{eqnarray}
L_n&=&\sum_{k=1}^{\infty}(n+k)p_k\frac{\partial}{\partial p_{n+k}}+\sum_{k=1}^{\infty}k(n-k)\frac{\partial^2}
{\partial p_k\partial p_{n-k}}+2Nn\frac{\partial}{\partial p_n}\nonumber\\
&&+N^2 \delta_{n,0}+Np_1\delta_{n+1,0}-(n+2)\frac{\partial}{\partial p_{n+2}}.
\end{eqnarray}

The character expansion of the Gaussian hermitian one-matrix model can be derived recursively
from a single $w$-constraint \cite{Mironov2105}
\begin{eqnarray}\label{CEGHM}
(\hat W_{-2}-l_0\{p\})Z_{G}\{p\}=0,
\end{eqnarray}
where $l_0\{p\}=\sum_{k=1}^{\infty}kp_k\frac{\partial}{\partial p_k}$.

Note that there is an intrinsic connection between (\ref{CEGHM}) and the Virasoro constraints (\ref{VCGHM}).
More precisely,  (\ref{CEGHM}) is equivalent to the sum of Virasoro constraints
\begin{equation}\label{VGcon}
\sum_{n=1}^{\infty} p_n L_{n-2}Z_G\{p\}=0.
\end{equation}

We see that the expressions of (\ref{l0wk}) and (\ref{l0vk}) are similar with (\ref{CEGHM}).
However, unlike the case of (\ref{CEGHM}), we can not give the (super) Virasoro constraints for (\ref{eznm})
from (\ref{l0wk}) and (\ref{l0vk}).

To present the desired constraints, let us introduce the operator $\hat l_0=l_0\{g\}+\tilde l_0\{\tilde g\}-l_0\{p\}-\tilde l_0\{\tilde p\}$,
where $\tilde l_0\{\tilde p\}=\sum_{k=1}^{\infty}k\tilde p_k\frac{\partial}{\partial \tilde p_k}$, such that it satisfies
\begin{eqnarray}
[\hat l_0, {\mathbb W}_{k}]=-k{\mathbb W}_{k},\quad
[\hat l_0, {\mathbb V}_{l}]=-l{\mathbb V}_{l}.
\end{eqnarray}

Thus we may further construct the constraints
\begin{eqnarray}\label{vircons1}
&&L^{\alpha_1,\cdots,\alpha_a}_{k_1,\cdots,k_{a+2a'}}\mathcal{Z}_{-n_1,-n_2}\{\vec u;p,\tilde p|g,\tilde g\}=0,\nonumber\\
&&G^{\alpha_1,\cdots,\alpha_a}_{k_1,\cdots,k_{a+2a'+1}}
\mathcal{Z}_{-n_1,-n_2}\{\vec u;p,\tilde p|g,\tilde g\}=0,
\end{eqnarray}
where
\begin{eqnarray}\label{conopn}
L^{\alpha_1,\cdots,\alpha_a}_{k_1,\cdots,k_{a+2a'}}=\hat l_0 \prod_{i=1}^a
{\mathbb W}_{k_i}^{\alpha_i}\prod_{j=1}^{2a'}{\mathbb V}_{k_{a+j}},\quad
G^{\alpha_1,\cdots,\alpha_a}_{k_1,\cdots,k_{a+2a'+1}}=\hat l_0 \prod_{i=1}^a
{\mathbb W}_{k_i}^{\alpha_i}\prod_{j=1}^{2a'+1}{\mathbb V}_{k_{a+j}},
\end{eqnarray}
$k_i,\alpha_i\in \mathbb{N}_+$ $(i=1,\cdots, a)$, $k_{a+j}\in \mathbb{N}$ $(j=1,\cdots, 2a'+1)$ and $a'\in \mathbb{N}$,
all lower indices of non-zero ${\mathbb V}$-operators have to be unequal.

It is interesting to note that the constraint operators (\ref{conopn}) yield the new infinite-dimensional
super algebra
\begin{align}\label{gsvir}
[L^{\alpha_1,\cdots,\alpha_a}_{k_1,\cdots,k_{a+2a'}}, L^{\beta_1,\cdots,\beta_b}_{l_1,\cdots,l_{b+2b'}}]&=
(\sum_{i=1}^a k_i\alpha_i+\sum_{i=1}^{2a'}k_{a+i}-\sum_{j=1}^b l_j\beta_j-\sum_{j=1}^{2b'}l_{b+j})\nonumber\\
&\ \ \ \ \cdot L^{\alpha_1,\cdots,\alpha_a,\beta_1,\cdots,\beta_b}_{k_1,\cdots,k_a,l_1,\cdots,l_b,
k_{a+1},\cdots,k_{a+2a'},l_{b+1},\cdots,l_{b+2b'}},\nonumber\\
[L^{\alpha_1,\cdots,\alpha_a}_{k_1,\cdots,k_{a+2a'}}, G^{\beta_1,\cdots,\beta_b}_{l_1,\cdots,l_{b+2b'+1}}]&=
(\sum_{i=1}^a k_i\alpha_i+\sum_{i=1}^{2a'}k_{a+i}-\sum_{j=1}^b l_j\beta_j-\sum_{j=1}^{2b'+1}l_{b+j})\nonumber\\
&\ \ \ \ \cdot G^{\alpha_1,\cdots,\alpha_a,\beta_1,\cdots,\beta_b}_{k_1,\cdots,k_a,l_1,\cdots,l_b,
k_{a+1},\cdots,k_{a+2a'},l_{b+1},\cdots,l_{b+2b'+1}},\nonumber\\
\{G^{\alpha_1,\cdots,\alpha_a}_{k_1,\cdots,k_{a+2a'+1}}, G^{\beta_1,\cdots,\beta_b}_{l_1,\cdots,l_{b+2b'+1}}\}
&=G^{\alpha_1,\cdots,\alpha_a}_{k_1,\cdots,k_{a+2a'+1}}G^{\beta_1,\cdots,\beta_b}_{l_1,\cdots,l_{b+2b'+1}}
+G^{\beta_1,\cdots,\beta_b}_{l_1,\cdots,l_{b+2b'+1}}G^{\alpha_1,\cdots,\alpha_a}_{k_1,\cdots,k_{a+2a'+1}}\nonumber\\
&=(\sum_{i=1}^a k_i\alpha_i+\sum_{i=1}^{2a'+1}k_{a+i}-\sum_{j=1}^b l_j\beta_j-\sum_{j=1}^{2b'+1}l_{b+j})\nonumber\\
&\ \ \ \ \cdot L^{\alpha_1,\cdots,\alpha_a,\beta_1,\cdots,\beta_b}_{k_1,\cdots,k_a,l_1,\cdots,l_b,
k_{a+1},\cdots,k_{a+2a'+1},l_{b+1},\cdots,l_{b+2b'+1}},
\end{align}
and the null super $3$-algebra
\begin{eqnarray}\label{null3alg}
&&[L^{\alpha_1,\cdots,\alpha_a}_{k_1,\cdots,k_{a+2a'}}, L^{\beta_1,\cdots,\beta_b}_{l_1,\cdots,l_{b+2b'}},
L^{\gamma_1,\cdots,\gamma_c}_{j_1,\cdots,j_{c+2c'}}]=0,\nonumber\\
&&[L^{\alpha_1,\cdots,\alpha_a}_{k_1,\cdots,k_{a+2a'}}, L^{\beta_1,\cdots,\beta_b}_{l_1,\cdots,l_{b+2b'}},
G^{\gamma_1,\cdots,\gamma_c}_{j_1,\cdots,j_{c+2c'+1}}]=0,\nonumber\\
&&[L^{\alpha_1,\cdots,\alpha_a}_{k_1,\cdots,k_{a+2a'}}, G^{\beta_1,\cdots,\beta_b}_{l_1,\cdots,l_{b+2b'+1}},
G^{\gamma_1,\cdots,\gamma_c}_{j_1,\cdots,j_{c+2c'+1}}]=0,\nonumber\\
&&[G^{\alpha_1,\cdots,\alpha_a}_{k_1,\cdots,k_{a+2a'+1}}, G^{\beta_1,\cdots,\beta_b}_{l_1,\cdots,l_{b+2b'+1}},
G^{\gamma_1,\cdots,\gamma_c}_{j_1,\cdots,j_{c+2c'+1}}]=0,
\end{eqnarray}
where the super $3$-bracket is defined by \cite{Hanlon}
\begin{equation}\label{NDF}
[A_{1}, A_{2},A_{3}]=\epsilon _{123}^{i_{1}i_{2}
i_{3}}(-1)^{\sum_{j=1}^{2}|A_{i_j}|(\sum_{l=j+1 , i_l<i_j}^{3}|A_{i_l}|)}
 A_{i_{1}}A_{i_{2}}A_{i_{3}},
\end{equation}
in which $|A_{i_j}|$ is the parity of the superoperator $A_{i_j}$.

When particularized to the constraint operators $L^{\alpha}_{1}=\hat l_0
{\mathbb W}_{1}^{\alpha}$ and $G^{\alpha}_{1,1}=\hat l_0
{\mathbb W}_{1}^{\alpha}{\mathbb V}_{1}$, we have
\begin{align}
&[L^{\alpha}_{1}, L^{\beta}_{1}]=(\alpha-\beta)L^{\alpha+\beta}_{1},\label{scvir}\nonumber\\
&[L^{\alpha}_{1}, G^{\beta}_{1,1}]=(\alpha-\beta-1)G^{\alpha+\beta}_{1,1},\nonumber\\
&\{G^{\alpha}_{1,1}, G^{\beta}_{1,1}\}=0.
\end{align}
Since (\ref{scvir}) is the super Virasoro algebra, we call the infinite-dimensional super algebra (\ref{gsvir})
the generalized super Virasoro algebra.

\subsection{The positive branch of hierarchy}

Let us turn to construct the bosonic operators
\begin{equation}\label{ewkn}
{\mathcal{W}}_{k}^{(n_1,n_2)}(\vec{u})=
\left\{
  \begin{array}{ll}
   (-1)^{n_1+n_2}\beta^{-1}\prod_{i=n_1+1}^{n_1+n_2}{\rm ad}_{{\mathcal{W}}^*(u_i)}
\prod_{j=1}^{n_1}{\rm ad}_{{\mathcal{W}}^\circledast(u_j)}\frac{\partial}{\partial p_1} , & \hbox{$k=1$;} \\
\\
    \frac{(-1)^{k-1}}{(k-1)!}{\rm ad}_{{\mathcal{W}}_{1}^{(n_1,n_2+1)}(\vec{u})}
{{\mathcal{W}}_{1}^{(n_1,n_2)}(\vec{u})}, & \hbox{$k\geq 2$,}
  \end{array}
\right.
\end{equation}
and fermionic operators
\begin{equation}\label{evkn}
\mathcal{V}_{k}^{(n_1,n_2)}(\vec{u})=
\left\{
  \begin{array}{ll}
(-1)^{n_1+n_2}\beta^{-1}\prod_{i=n_1+1}^{n_1+n_2}{\rm ad}_{{\mathcal{W}}^*(u_i)}
\prod_{j=1}^{n_1}{\rm ad}_{{\mathcal{W}}^\circledast(u_j)}\frac{\partial}{\partial \tilde p_k}, & \hbox{$k=0,1$;} \\
\\
    \frac{1}{k-1}
[\mathcal{W}_{k-1}^{(n_1,n_2)}(\vec{u}),
\mathcal{V}_{1}^{(n_1,n_2+1)}(\vec{u})], & \hbox{$k\geq 2$,}
  \end{array}
\right.
\end{equation}
where $(n_1,n_2)\in \mathbb{N}^2\setminus (0,0)$.

There are the actions
\begin{subequations}
\begin{eqnarray}
\mathcal W_{k}^{(n_1,n_2)}(\vec{u})J_{\Lambda}&=&\sum_{\Gamma}C(n_1,n_2;\Lambda,\Gamma)
\frac{\langle \beta^{-1}k\frac{\partial}{\partial p_k}J_{\Lambda} ,J_{\Gamma}\rangle_{\beta}}
{\langle J_{\Gamma},J_{\Gamma}\rangle_{\beta}}J_{\Gamma},\label{wkn1n2act}\\
\mathcal V_{k}^{(n_1,n_2)}(\vec{u})J_{\Lambda}&=&\sum_{\Gamma}C(n_1,n_2;\Lambda,\Gamma)
\frac{\langle \beta^{-1}\frac{\partial}{\partial \tilde p_k}J_{\Lambda} ,J_{\Gamma}\rangle_{\beta}}
{\langle J_{\Gamma},J_{\Gamma}\rangle_{\beta}}J_{\Gamma},\label{vkn1n2act}
\end{eqnarray}
\end{subequations}
where
the sum is over the superpartitions $\Gamma$ satisfying $\Gamma\subseteq \Lambda$, $|\Lambda^*/\Gamma^*|=k$,  $|\Lambda^\circledast/\Gamma^\circledast|=k$
in (\ref{wkn1n2act}) and $|\Lambda^\circledast/\Gamma^\circledast|=k+1$ in (\ref{vkn1n2act}).

We note that in terms of variables $x_i$ and $\theta_i$,
${\mathcal W}_{2}^{(1,0)}(0)$ is given by
\begin{eqnarray}
{\mathcal W}_{2}^{(1,0)}(0)
=\sum_{i=1}^{N}\frac{\partial^2}{\partial x^2_i}+2\beta\sum_{1\leq i\neq j\leq N}\frac{1}{x_i-x_j}\frac{\partial}{\partial x_i}
-2\beta\sum_{1\leq i\neq j\leq N}\frac{\theta_i-\theta_j}{(x_i-x_j)^2}\frac{\partial}{\partial \theta_i}.
\end{eqnarray}
It is the gauged Hamiltonian (with $\omega=0$) for the supersymmetric rational CMS model \cite{Desrosiers03}.

By means of the operators (\ref{ewkn}) and (\ref{evkn}), we give the positive branch of supersymmetric $\beta$-deformed hierarchy
\begin{eqnarray}\label{eszngp}
{\mathcal Z}_{n_1,n_2}\{\vec u;p,\tilde p|g,\tilde g|h,\tilde h\}
&=&e^{\beta\sum_{k=1}^{\infty}({\mathcal{W}}_{k}^{(n_1,n_2)}(\vec{u})\frac{h _k}{k}+{\mathcal{V}}_{k-1}^{(n_1,n_2)}
(\vec{u})\tilde h_{k-1})}e^{\beta\sum_{k=1}^{\infty}
(\frac{p_kg_k}{k}+\tilde p_{k-1} \tilde g_{k-1})}\nonumber\\
&=&\sum_{\lambda,\mu}
\prod_{r=1}^{n_1+n_2}\frac{E_{u_r}(J_{\lambda})E'_{\beta}(J_{\mu})}
{E_{u_r}(J_{\mu})E'_{\beta}(J_{\lambda})}
\frac{J_{\lambda/\mu}\{h\}}
{\langle J_{\lambda}, J_{\lambda}\rangle_{\beta}}J_{\mu}\{p\}J_{\lambda}\{g\}
+\sum_{\Lambda,\mu}\frac{E'_{\beta}(J_{\mu})^{n_1+n_2}}{E'_\beta(J_{\Lambda})^{n_1+n_2}}
\nonumber\\
&&\cdot \prod_{r_1=1}^{n_1}\prod_{r_2=n_1+1}^{n_1+n_2}
\frac{E_{u_{r_1}}(J_{\Lambda})E_{u_{r_1}}(J_{\delta_{m+1}})\tilde E_{u_{r_2}}(J_{\Lambda})E_{u_{r_2}}(J_{\delta_{m}})}
{E_{u_{r_1}}(J_{\mu})E'_{\beta}(J_{\delta_{m+1}})E_{u_{r_2}}(J_{\mu})E'_{\beta}(J_{\delta_{m}})}
J_{\Lambda/\mu}\{h,\tilde h\}
\nonumber\\
&&\cdot J_{\mu}\{p\}\frac{J_{\Lambda}\{g,\tilde g\}}
{\langle J_{\Lambda}, J_{\Lambda}\rangle_{\beta}}+\sum_{\Lambda,\Gamma}
\frac{E'_{\beta}(J_{\Gamma})^{n_1+n_2}}{E'_\beta(J_{\Lambda})^{n_1+n_2}}
\prod_{r_1=1}^{n_1}\prod_{r_2=n_1+1}^{n_1+n_2}\frac{E_{u_{r_1}}(J_{\Lambda})}
{E_{u_{r_1}}(J_{\Gamma})}\nonumber\\
&&\cdot \frac{E_{u_{r_1}}(J_{\delta_{m+1}})E'_{\beta}(J_{\delta_{m'+1}})\tilde E_{u_{r_2}}(J_{\Lambda})E_{u_{r_2}}(J_{\delta_{m}})E'_{\beta}(J_{\delta_{m'}})}
{E_{u_{r_1}}(J_{\delta_{m'+1}})E'_{\beta}(J_{\delta_{m+1}})\tilde E_{u_{r_2}}(J_{\Gamma})E_{u_{r_2}}(J_{\delta_{m'}})E'_{\beta}(J_{\delta_{m}})}
\frac{J_{\Lambda/\Gamma}
\{h,\tilde h\}}{\langle J_{\Lambda}, J_{\Lambda}\rangle_{\beta}}
\nonumber\\
&&\cdot J_{\Gamma}\{p,\tilde p\}J_{\Lambda}\{g,\tilde g\},
\end{eqnarray}
where $\lambda$ and $\mu$ are regular partitions, $\Lambda$ and $\Gamma$ are superpartitions of fermionic
degree $m>0$ and $m'>0$, respectively.

There are the constraints for (\ref{eszngp})
\begin{eqnarray}\label{hcons}
&&\bar{\mathbb{W}}_k{\mathcal Z}_{n_1,n_2}
\{\vec u;p,\tilde p|g,\tilde g|h,\tilde h\}=0,\nonumber\\
&&\bar{\mathbb{V}}_l{\mathcal Z}_{n_1,n_2}\{\vec u;p,\tilde p|g,\tilde g|h,\tilde h\}=0,
\end{eqnarray}
where the constraint operators are given by
\begin{eqnarray}
&&\bar{\mathbb{W}}_k=\beta^{-1}k\frac{\partial}{\partial h_k}-{\mathcal{W}}_{k}^{(n_1,n_2)}(\vec{u}), \ \ k\geq 1, \nonumber\\
&&\bar{\mathbb{V}}_l=\beta^{-1}\frac{\partial}{\partial \tilde h_l}- {\mathcal{V}}_{l}^{(n_1,n_2)}(\vec{u}), \ \ l\geq 0,
\end{eqnarray}
they also yield null super algebra.

Let $\bar l_0=l_0\{h\}+\tilde l_0\{\tilde h\}+l_0\{p\}+\tilde l_0\{\tilde p\}$, we then have
\begin{eqnarray}
[\bar l_0, \bar{\mathbb{W}}_k]=-k\bar{\mathbb{W}}_k,\quad
[\bar l_0, \bar{\mathbb{V}}_l]=-l\bar{\mathbb{V}}_l.
\end{eqnarray}

Thus we may construct the generalized super Virasoro constraints
\begin{eqnarray}\label{vircons2}
&&\bar L^{\alpha_1,\cdots,\alpha_a}_{k_1,\cdots,k_{a+2a'}}{\mathcal Z}_{n_1,n_2}\{\vec u;p,\tilde p|g,\tilde g|h,\tilde h\}=0,\nonumber\\
&&\bar G^{\alpha_1,\cdots,\alpha_a}_{k_1,\cdots,k_{a+2a'+1}}{\mathcal Z}_{n_1,n_2}\{\vec u;p,\tilde p|g,\tilde g|h,\tilde h\}=0,
\end{eqnarray}
where
\begin{eqnarray}\label{conopp}
\bar L^{\alpha_1,\cdots,\alpha_a}_{k_1,\cdots,k_{a+2a'}}=\bar l_0 \prod_{i=1}^a
\bar{\mathbb W}_{k_i}^{\alpha_i}\prod_{j=1}^{2a'}\bar{\mathbb V}_{k_{a+j}},\quad
\bar G^{\alpha_1,\cdots,\alpha_a}_{k_1,\cdots,k_{a+2a'+1}}=\bar l_0 \prod_{i=1}^a
\bar{\mathbb W}_{k_i}^{\alpha_i}\prod_{j=1}^{2a'+1}\bar{\mathbb V}_{k_{a+j}},
\end{eqnarray}
$k_i,\alpha_i\in \mathbb{N}_+$ $(i=1,\cdots, a)$, $k_{a+j}\in \mathbb{N}$ $(j=1,\cdots, 2a'+1)$ and $a'\in \mathbb{N}$.
These constraint operators satisfy the generalized super Virasoro algebra (\ref{gsvir}) and null super $3$-algebra (\ref{null3alg}).

We have given the desired supersymmetric $\beta$-deformed hierarchy.
The non-deformed hierarchy follows from the $\beta=1$ case and the Jack superpolynomials in the character expansions are replaced
by the Schur-Jack superpolynomials.
Setting to zero all the fermionic variables in the supersymmetric $\beta$-deformed hierarchy, we arrive at the results in Refs. \cite{2301.11877,2301.12763}
\begin{subequations}\label{betaZ}
\begin{eqnarray}
\bar{Z}_{-n}\{\vec u;p,\bar{p}\}
&=&\sum_{\lambda}\prod_{i=1}^n\frac{J_{\lambda}\{p_k=u_i\}}{J_{\lambda}
\{p_k=\beta^{-1}\delta_{k,1}\}}
\frac{J_{\lambda}\{p\}J_{\lambda}\{\bar{p}\}}{\langle J_{\lambda},J_{\lambda}\rangle_{\beta}},\label{betaz-n}\\
\bar{Z}_{n} \{\vec u;p,\bar{p},g\}&=&\sum_{\lambda,\mu}\prod_{i=1}^n\frac{J_\lambda\{p_k=u_i\}
J_\mu\{p_k=\beta^{-1}\delta_{k,1}\}}
{J_\lambda\{p_k=\beta^{-1}\delta_{k,1}\}J_\mu\{p_k=u_i\}}
\frac{J_{\lambda/\mu}\{\bar p\}J_\mu\{p\}J_\lambda\{g\}}
{\langle J_\lambda,J_\lambda\rangle_{\beta}}.\label{betazn}
\end{eqnarray}
\end{subequations}
It is noted that the $\beta$-deformed eigenvalue models in Refs. \cite{Morozov,Cassia} are special cases of (\ref{betaz-n}).
Furthermore, the (skew) hypergeometric Hurwitz $\tau$-functions \cite{Al,2301.04107,2301.11877} correspond to the
$\beta=1$ case in (\ref{betaZ}).

%%%%%%%%%%%%%%%%%%%%%%%%%%%%%%%%%%%%%%%%%%%%%%%%%%%%%%%%%%%%%%%%%%%%%%%%%%%%%%%%%%%%%%%%%%%%%
\section{Supersymmetric $(q,t)$-deformed partition function hierarchy}
%%%%%%%%%%%%%%%%%%%%%%%%%%%%%%%%%%%%%%%%%%%%%%%%%%%%%%%%%%%%%%%%%%%%%%%%%%%%%%%%%%%%%%%%%%%%%
\subsection{Supersymmetric $(q,t)$-deformed HK partition functions}
The Macdonald superpolynomials $P_{\Lambda}$ are eigenfunctions of two family of commuting superoperators \cite{1202.3922}
\begin{eqnarray}
D^*(z;q,t)&=&\sum_{m=0}^N\sum_{\sigma\in S_N/(S_m\times S_{m^c})}
{\mathcal K}_{\sigma}\left(\frac{\Delta_m}{\Delta^t_m}\prod_{i=1}^N
(1+zY_i)\frac{\Delta^t_m}{\Delta_m}\pi_{1,\cdots,m}
\right),\nonumber\\
D^\circledast(z;q,t)&=&\sum_{m=0}^N\sum_{\sigma\in S_N/(S_m\times S_{m^c})}
{\mathcal K}_{\sigma}\left(\frac{\Delta_m}{\Delta^t_m}\prod_{i=1}^m
(1+zqY_i)\prod_{i=m+1}^N
(1+zY_i)\frac{\Delta^t_m}{\Delta_m}\pi_{1,\cdots,m}
\right),
\end{eqnarray}
where $S_m$ and $S_{m^c}$ are groups of permutations of $x_1,\cdots,x_m$ and $x_{m+1},\cdots,x_{N}$, respectively,
${\mathcal K}_{\sigma}=\kappa_\sigma K_\sigma$ with
$K_\sigma: (x_1,\cdots,x_N)\mapsto (x_{\sigma_1},\cdots,x_{\sigma_N})$ and
$\kappa_\sigma: (\theta_1,\cdots,\theta_N)\mapsto (\theta_{\sigma_1},\cdots,\theta_{\sigma_N})$,
$\Delta^t_m=\prod_{1\leq i<j\leq N}(tx_i-x_j)$,
$Y_i$ are Cherednick operators, $\pi_{1,\cdots,m}=\prod_{i=1}^{m}\theta_i
\frac{\partial}{\partial \theta_i}\prod_{j=m+1}^{N}
\frac{\partial}{\partial \theta_j}\theta_j$.

There are the actions
\begin{eqnarray}\label{Dstaract}
D^{\circ}(z;q,t)P_{\Lambda}=\mathcal{E}_{\Lambda^\circ}(z;q,t)P_{\Lambda},
\end{eqnarray}
where $\mathcal{E}_{\lambda}(z;q,t)=\prod_{i=1}^N(1+zq^{\lambda_i}t^{1-i})$.

Let $D^{\circ}(z;q,t)=\sum_{n=1}^{N}D^{\circ}_nz^n$.
In terms of the supercharges $Q_3=\sum_{i}A_i(t)\xi_i\tau_i\frac{\partial }{\partial \theta_i}$ and $Q_4=\sum_{i}\theta_i$
in the integrable supersymmetric trigonometric Ruijsenaars-Schneider model,
$D^\circ_1$ are represented as \cite{Blondeau2015}
\begin{eqnarray}
D^*_1&=&t^{1-N}\{Q_3,Q_4\}, \nonumber\\
D^\circledast_1&=&t^{1-N}(q-1)(t-1)\sum_{i,j}\frac{x_iA_i(t)}{tx_i-x_j}
\theta_j\frac{\partial }{\partial \theta_i}\xi_i\tau_i+D^*_1,
\end{eqnarray}
where $A_i(t)=\prod_{i\neq j}\frac{tx_i-x_j}{x_i-x_j}$,
$\xi_i=\sum_{I\subset \{1,2,\cdots,N\}}\prod_{j\in I,j\neq i}\frac{(qtx_i-x_j)(x_i-x_j)}{(qx_i-x_j)(tx_i-x_j)}\theta_{I}\varrho_{I}$,
$\varrho_{I}$ picks up the coefficient of $\theta_{I}$ in a superpolynomial,
i.e., $\varrho_{I}\theta_{J}=\delta_{IJ}$, $\tau_i$ is the $q$-shift operator such that $x_i\mapsto qx_i$ and $x_j\mapsto x_j$ for $i\neq j$.

Let us introduce the $(q,t)$-deformed Hurwitz superoperators
\begin{eqnarray}\label{qtsope}
\mathcal{\hat W}^{\circ}(u)=\frac{1}{1-q}(E+\delta\bar E)+\frac{t^{u+1}}{(1-q)^2}
(t^{-1}D^\circ_1+\frac{1-t^{-N}}{1-t}).
\end{eqnarray}
They act on the Macdonald superpolynomials $P_{\Lambda}$ as
\begin{eqnarray}\label{w0act-mac}
\mathcal{\hat W}^{\circ}(u)P_{\Lambda}=\sum_{(i,j)\in \Lambda^{\circ}}{\hat c}(u;i,j)P_{\Lambda},
\end{eqnarray}
where ${\hat c}(u;i,j)=\frac{1-q^{j-1}t^{u+1-i}}{1-q}$.

The supersymmetric $(q,t)$-deformed HK partition functions can be generated by the superoperators (\ref{qtsope})
\begin{eqnarray}\label{2qtSHKM}
{\mathcal {\hat Z}}^{\circ}\{p,\tilde p\}
&=&e^{\tau\mathcal{\hat W}^{\circ}(N)} e^{\frac{1-t}{1-q}e^{-\tau N}(p_1+\tilde p_1\theta)}\nonumber\\
&=&\sum_{\Lambda}e^{\tau\sum_{(i,j)\in \Lambda^{\circ}}{\hat c}(N;i,j)}P_{\Lambda}\{p,\tilde p\}Q_{\Lambda}
\{g_k=e^{-\tau N}\delta_{k,1},\tilde g_k=\frac{1-t}{q^{-1}-1}e^{-\tau N}\theta\delta_{k,1}\}.
\end{eqnarray}
Taking the limit $t=q^{\beta}, q\rightarrow 1$ in (\ref{2qtSHKM}), it reduces to the $\beta$-deformed models
(\ref{SHKM}).

Here for the character expansions in (\ref{2qtSHKM}), we have used the actions (\ref{w0act-mac}) and Cauchy formula
\begin{equation}
e^{\sum_{k=1}^{\infty}(\frac{1-t^k}{1-q^k}\frac{p_kg_k}{k}+q^{1-k}\tilde p_{k-1} \tilde g_{k-1})}=\sum_{\Lambda}
P_{\Lambda}\{p,\tilde p\}Q_{\Lambda}\{g,\tilde g\},
\end{equation}
where $Q_{\Lambda}\{g,\tilde g\}=\langle P_{\Lambda},P_{\Lambda}\rangle_{q,t}^{-1}P_{\Lambda}\{g,\tilde g\}$,
the $(q,t)$-deformed scalar product is given by $\langle P_{\Lambda},P_{\Lambda}\rangle_{q,t}=(-1)^{\frac{m(m-1)}{2}}q^{|\Lambda^a|}
\frac{{\bar h}^{up}_{\Lambda}}{{\bar h}^{lo}_{\Lambda}}$, $m$ is the fermionic degree of $\Lambda$,
${\bar h}^{up}_{\Lambda}=\prod_{s\in {\mathcal B}\Lambda}(1-q^{a_{\Lambda^*}(s)+1}t^{l_{\Lambda^\circledast}(s)})$,
${\bar h}^{lo}_{\Lambda}=\prod_{s\in {\mathcal B}\Lambda}(1-q^{a_{\Lambda^\circledast}(s)}t^{l_{\Lambda^*}(s)+1})$.

%%%%%%%%%%%%%%%%%%%%%%%%%%%%%%%%%%%%%%%%%%%%%%%%%%%%%%%%%%%%%%%%%%%%%%%%%%%%%%%%%%%%%%%%%%%%%}
\subsection{Supersymmetric $(q,t)$-deformed partition function hierarchy}
%%%%%%%%%%%%%%%%%%%%%%%%%%%%%%%%%%%%%%%%%%%%%%%%%%%%%%%%%%%%%%%%%%%%%%%%%%%%%%%%%%%%%%%%%%%%%
Let us introduce the superoperators
\begin{eqnarray}\label{O1}
O^\circ(u;q,t)=(1-q)^{-(E+\delta\bar E)}\prod_{k=1}^{\infty}\frac{D^\circ(-q^{-k}t^u;q,t)}
{\prod_{i=1}^N(1-q^{-k}t^{u+1-i})},
\end{eqnarray}
where $|q|>1$.

Then by the actions (\ref{Dstaract}), we have
\begin{eqnarray}
O^{\circ}(u;q,t)P_\Lambda=\prod_{(i,j)\in \Lambda^{\circ}}{\hat c}(u;i,j)P_\Lambda.
\end{eqnarray}

We further define the operators
\begin{eqnarray}\label{e4opes}
\hat{\mathcal{W}}^{(n_1,n_2)}_{-k}(\vec u)&=&O(n_1,n_2;q,t)p_k(O(n_1,n_2;q,t))^{-1},\nonumber\\
\hat{\mathcal{W}}^{(n_1,n_2)}_{k}(\vec u)&=&k\frac{1-q^k}{1-t^k}(O(n_1,n_2;q,t))^{-1}
\frac{\partial}{\partial p_k}O(n_1,n_2;q,t),\ \ k>0,\nonumber\\
\hat{\mathcal{V}}^{(n_1,n_2)}_{-l}(\vec u)&=&O(n_1,n_2;q,t)
\tilde p_l(O(n_1,n_2;q,t))^{-1},\nonumber\\
\hat{\mathcal{V}}^{(n_1,n_2)}_{l}(\vec u)&=&\frac{1-q^l}{1-t^l}(O(n_1,n_2;q,t))^{-1}
\frac{\partial}{\partial \tilde p_l}O(n_1,n_2;q,t),\ \ l\geq 0,
\end{eqnarray}
where $O(n_1,n_2;q,t)=\prod_{r_1=1}^{n_1}O^\circledast(u_{r_1};q,t)
\prod_{r_2=n_1+1}^{n_1+n_2}O^*(u_{r_2};q,t)$, $(n_1,n_2)\in \mathbb{N}^2\setminus (0,0)$.

There are the actions
\begin{subequations}
\begin{eqnarray}
\hat{\mathcal{W}}^{(n_1,n_2)}_{-k}(\vec u)P_\Lambda
&=&\sum_{\Omega}\hat C(n_1,n_2;\Omega,\Lambda)\langle p_kP_{\Lambda},Q_{\Omega}\rangle_{q,t}
 P_{\Omega},\label{w-kk-mac}\\
\hat{\mathcal{V}}^{(n_1,n_2)}_{-k}(\vec u)P_\Lambda
&=&\sum_{\Omega}\hat C(n_1,n_2;\Omega,\Lambda)\langle \tilde p_kP_{\Lambda},Q_{\Omega}\rangle_{q,t}
 P_{\Omega},\label{v-kk-mac}
\end{eqnarray}
\end{subequations}
where $\hat C(n_1,n_2;\Omega,\Lambda)=\prod_{(i_1,j_1)\in\Omega^\circledast/\Lambda^\circledast}
\prod_{r_1=1}^{n_1}\hat c(u_{r_1};i_1,j_1)
\prod_{(i_2,j_2)\in\Omega^*/\Lambda^*}
\prod_{r_2=n_1+1}^{n_1+n_2}\hat c(u_{r_2};i_2,j_2)$,
the sum is over the superpartitions $\Omega$ satisfying $\Lambda \subseteq \Omega$, $|\Omega^*/\Lambda^*|=k$,
$|\Omega^\circledast/\Lambda^\circledast|=k$ in (\ref{w-kk-mac}) and $|\Omega^\circledast/\Lambda^\circledast|=k+1$ in
(\ref{v-kk-mac});\\
and
\begin{subequations}
\begin{eqnarray}
\hat{\mathcal W}_{k}^{(n_1,n_2)}(\vec{u})P_{\Lambda}&=&\sum_{\Gamma}
\hat C(n_1,n_2;\Lambda,\Gamma)
\langle k\frac{1-q^k}{1-t^k}\frac{\partial}{\partial p_k}P_{\Lambda},Q_{\Gamma}\rangle_{q,t}
P_{\Gamma},\label{wkk-mac}\\
\hat{\mathcal V}_{k}^{(n_1,n_2)}(\vec{u})P_{\Lambda}&=&\sum_{\Gamma}
\hat C(n_1,n_2;\Lambda,\Gamma)
\langle \frac{1-q^k}{1-t^k}\frac{\partial}{\partial \tilde p_k}P_{\Lambda},Q_{\Gamma}\rangle_{q,t}
P_{\Gamma},\label{vkk-mac}
\end{eqnarray}
\end{subequations}
where the sum is over the superpartitions $\Gamma$ satisfying $\Gamma\subseteq \Lambda$,
$|\Lambda^*/\Gamma^*|=k$, $|\Lambda^\circledast/\Gamma^\circledast|=k$ in (\ref{wkk-mac}) and
$|\Lambda^\circledast/\Gamma^\circledast|=k+1$ in (\ref{vkk-mac}).

We may give the negative and positive branches of supersymmetric $(q,t)$-deformed hierarchy, respectively,
\begin{subequations}\label{ehatzn}
\begin{eqnarray}
\hat{\mathcal{Z}}_{-n_1,-n_2}\{\vec u;p,\tilde p|g, \tilde g\}&=&
e^{\sum_{k=1}^{\infty}\frac{1-t^k}{1-q^k}(\hat{\mathcal{W}}_{-k}^{(n_1,n_2)}
(\vec{u})\frac{g_k}{k}+\hat{\mathcal{V}}_{-k+1}^{(n_1,n_2)}
(\vec{u})\tilde g_{k-1})}
\cdot 1\nonumber\\
&=&\sum_{\lambda}\prod_{r=1}^{n_1+n_2}
\frac{\mathcal{E}_{t^{u_r}}(P_{\lambda})}{\mathcal{E}'_{0}(P_{\lambda})}
P_{\lambda}\{p\}
Q_{\lambda}\{g\}+\sum_{\Lambda}\prod_{r_1=1}^{n_1}
\prod_{r_2=n_1+1}^{n_1+n_2}
\frac{\mathcal{E}_{t^{-m+u_{r_1}}}(P_{\Lambda})}{\mathcal{E}'_{0}(P_{\Lambda})
}\nonumber\\
&&\cdot \frac{\mathcal{E}_{t^{u_{r_1}}}(P_{\delta_{m+1}})\mathcal{\tilde E}_{t^{-m+1+u_{r_2}}}(P_{\Lambda})
\mathcal{E}_{t^{u_{r_2}}}(P_{\delta_{m}})}
{\mathcal{E}'_{0}(P_{\delta_{m+1}})\mathcal{\tilde E}'_{0}(P_{\Lambda})\mathcal{E}'_{0}(P_{\delta_{m}})}
P_{\Lambda}\{p,\tilde p\}Q_{\Lambda}\{g,\tilde g'\},\label{ehatzn1}\\
\hat{\mathcal Z}_{n_1,n_2}\{\vec u;p,\tilde p|g,\tilde g|h,\tilde h\}
&=&e^{\sum_{k=1}^{\infty}\frac{1-t^k}{1-q^k}(\hat{\mathcal{W}}_{k}^{(n_1,n_2)}
(\vec{u})
\frac{h_k}{k}+\hat{\mathcal{V}}_{k-1}^{(n_1,n_2)}
(\vec{u})\tilde h_{k-1})}e^{\sum_{k=1}^{\infty}\frac{1-t^k}{1-q^k}(\frac{p_kg_k}{k}
+\tilde p_{k-1} \tilde g_{k-1})}\nonumber\\
&=&\sum_{\lambda,\mu}
\prod_{r=1}^{n_1+n_2}\frac{\mathcal{E}_{t^{u_r}}(P_{\lambda})\mathcal{E}'_{0}(P_{\mu})}
{\mathcal{E}_{t^{u_r}}(P_{\mu})
\mathcal{E}'_{0}(P_{\lambda})}
P_{\lambda/\mu}\{h\}P_{\mu}\{p\}Q_{\lambda}\{g\}+\sum_{\Lambda,\mu}
(\mathcal{E}'_{0}(P_{\mu}))^{n_1+n_2}
\nonumber\\
&&\cdot \prod_{r_1=1}^{n_1}\prod_{r_2=n_1+1}^{n_1+n_2}
\frac{\mathcal{E}_{t^{-m+u_{r_1}}}(P_{\Lambda})
\mathcal{E}_{t^{u_{r_1}}}(P_{\delta_{m+1}})
\mathcal{\tilde E}_{t^{-m+1+u_{r_2}}}(P_{\Lambda})
\mathcal{E}_{t^{u_{r_2}}}(P_{\delta_m})}
{\mathcal{E}'_{0}(P_{\Lambda})\mathcal{E}_{t^{u_{r_1}}}(P_{\mu})
\mathcal{E}'_{0}(P_{\delta_{m+1}})\mathcal{\tilde E}'_{0}(P_{\Lambda})\mathcal{E}_{t^{u_{r_2}}}(P_{\mu})
\mathcal{E}'_{0}(P_{\delta_m})}\nonumber\\
&&\cdot P_{\Lambda/\mu}\{h,\tilde h'\}P_{\mu}\{p\}Q_{\Lambda}\{g,\tilde g'\}
+\sum_{\Lambda,\Gamma}\prod_{r_1=1}^{n_1}\prod_{r_2=n_1+1}^{n_1+n_2}
\frac{\mathcal{E}_{t^{-m+u_{r_1}}}
(P_{\Lambda})\mathcal{E}'_{0}(P_{\Gamma})}{\mathcal{E}_{t^{-m'+u_{r_1}}}
(P_{\Gamma})\mathcal{E}'_{0}
(P_{\Lambda})}
\nonumber\\
&&\cdot \frac{\mathcal{E}_{t^{u_{r_1}}}(P_{\delta_{m+1}})
\mathcal{E}'_{0}(P_{\delta_{m'+1}})\mathcal{\tilde E}_{t^{-m+1+u_{r_2}}}(P_{\Lambda})\mathcal{\tilde E}'_{0}(P_{\Gamma})\mathcal{E}_{t^{u_{r_2}}}(P_{\delta_{m}})
\mathcal{E}'_{0}(P_{\delta_{m'}})}
{\mathcal{E}_{t^{u_{r_1}}}(P_{\delta_{m'+1}})
\mathcal{E}'_{0}(P_{\delta_{m+1}})\mathcal{\tilde E}_{t^{-m'+1+u_{r_2}}}(P_{\Gamma})\mathcal{\tilde E}'_{0}(P_{\Lambda})\mathcal{E}_{t^{u_{r_2}}}(P_{\delta_{m'}})
\mathcal{E}'_{0}(P_{\delta_{m}})}
\nonumber\\
&&\cdot P_{\Lambda/\Gamma}\{h,\tilde h'\}P_{\Gamma}\{p,\tilde p\}Q_{\Lambda}\{g,\tilde g'\},\label{ehatzn2}
\end{eqnarray}
\end{subequations}
where $\lambda$ and $\mu$ are regular partitions, $\Lambda$ and $\Gamma$ are superpartitions of fermionic
degree $m>0$ and $m'>0$, respectively, $P_{\Lambda/\Gamma}=\sum_{\Omega}\langle  P_\Lambda, Q_\Gamma Q_\Omega \rangle_{q,t}P_\Omega$
is the skew Macdonald superpolynomial,
the scaled fermionic variables $\tilde g'_{l}$ and $\tilde h'_{l}$, $l=0,1,\cdots$ are
$\tilde g'_{l}=\frac{1-t^{l+1}}{q^{-l}-q}\tilde g_{l}$, $\tilde h'_{l}=\frac{1-t^{l+1}}{q^{-l}-q}\tilde h_{l}$.

Here we have used the evaluation formulas \cite{Lapointe2020}
\begin{eqnarray}\label{evm}
\frac{\mathcal{E}_{u}(P_{\Lambda})}{\mathcal{E}'_{0}(P_{\Lambda})}&=&
\prod_{(i,j)\in \Lambda^\circledast/\delta_{m+1}}\frac{1-q^{j-1}t^{m-i+1}u}{1-q},\ \ m\geq 0,\nonumber\\
\frac{\mathcal{\tilde E}_{u}(P_{\Lambda})}{\mathcal{\tilde E'}_{0}(P_{\Lambda})}&=&
\prod_{(i,j)\in \Lambda^*/\delta_m}\frac{1-q^{j-1}t^{m-i}u}{1-q},\ \ m\geq 1,
\end{eqnarray}
where $m$ is the fermionic
degree of $\Lambda$,
\begin{eqnarray}
\mathcal{E}'_{0}(P_{\Lambda})&=&(1-q)^{|\Lambda^\circledast/\delta_{m+1}|}
\mathcal{E}_{0}(P_{\Lambda})=\frac{(1-q)^{|\Lambda^\circledast/\delta_{m+1}|}t^{n(\Lambda^\circledast/\delta_{m+1})
+n((\Lambda')^a/\delta_m)}}{q^{(m-1)|\Lambda^a/\delta_m|
-n(\Lambda^a/\delta_m)}{\bar h}^{lo}_{\Lambda}},\nonumber\\
\mathcal{\tilde E'}_{0}(P_{\Lambda})&=&(1-q)^{|\Lambda^*/\delta_m|}\mathcal{\tilde E}_{0}(P_{\Lambda})=\frac{(1-q)^{|\Lambda^*/\delta_m|}t^{n(\Lambda^*/\delta_m)
+n((\Lambda')^a/\delta_{m-1})}}{q^{(m-2)|\Lambda^a/\delta_{m-1}|
-n(\Lambda^a/\delta_{m-1})}{\bar h}^{lo}_{\Lambda}},
\end{eqnarray}
and $n(\lambda)=\sum_{i}(i-1)\lambda_i$. When $m=0$ in (\ref{evm}), it gives the evaluation formula for the
Macdonald polynomial $P_{\lambda}$ associated
with the regular partition $\lambda$ \cite{Macdonaldbook}
\begin{eqnarray}
\frac{\mathcal{E}_{u}(P_{\lambda})}{\mathcal{E}'_{0}(P_{\lambda})}
=\frac{P_{\lambda}\{p_k=\frac{1-u^k}{1-t^k}\}}
{P_{\lambda}\{p_k=\frac{(1-q)^k}{1-t^k}\}}=\prod_{(i,j)\in\lambda}\frac{1-q^{j-1}t^{-i+1}u}{1-q}.
\end{eqnarray}

For the supersymmetric $(q,t)$-deformed partition functions (\ref{ehatzn}), there are the generalized super Virasoro constraints
\begin{eqnarray}\label{vircons3}
&&\hat L^{\alpha_1,\cdots,\alpha_a}_{k_1,\cdots,k_{a+2a'}}\mathcal{\hat Z}_{-n_1,-n_2}\{\vec u;p,\tilde p|g,\tilde g\}=0,\nonumber\\
&&\hat G^{\alpha_1,\cdots,\alpha_a}_{k_1,\cdots,k_{a+2a'+1}}\mathcal{\hat Z}_{-n_1,-n_2}\{\vec u;p,\tilde p|g,\tilde g\}=0,
\end{eqnarray}
and
\begin{eqnarray}\label{vircons4}
&&\tilde L^{\alpha_1,\cdots,\alpha_a}_{k_1,\cdots,k_{a+2a'}}{\mathcal {\hat Z}}_{n_1,n_2}\{\vec u;p,\tilde p|g,\tilde g|h,\tilde h\}=0,\nonumber\\
&&\tilde G^{\alpha_1,\cdots,\alpha_a}_{k_1,\cdots,k_{a+2a'+1}}{\mathcal {\hat Z}}_{n_1,n_2}\{\vec u;p,\tilde p|g,\tilde g|h,\tilde h\}=0,
\end{eqnarray}
where the constraint operators are given by
\begin{eqnarray}\label{dconopn}
&&\hat L^{\alpha_1,\cdots,\alpha_a}_{k_1,\cdots,k_{a+2a'}}=\hat l_0 \prod_{i=1}^a
\hat{\mathbb{W}}^{\alpha_i}_{k_i}\prod_{j=1}^{2a'}\hat{\mathbb V}_{k_{a+j}},\quad
\hat G^{\alpha_1,\cdots,\alpha_a}_{k_1,\cdots,k_{a+2a'+1}}=\hat l_0 \prod_{i=1}^a
\hat{\mathbb{W}}^{\alpha_i}_{k_i}\prod_{j=1}^{2a'+1}\hat{\mathbb V}_{k_{a+j}},
\end{eqnarray}
and
\begin{eqnarray}\label{dconopp}
&&\tilde L^{\alpha_1,\cdots,\alpha_a}_{k_1,\cdots,k_{a+2a'}}=\bar l_0 \prod_{i=1}^a
\tilde{\mathbb{W}}^{\alpha_i}_{k_i}\prod_{j=1}^{2a'}\tilde{\mathbb V}_{k_{a+j}},\quad
\tilde G^{\alpha_1,\cdots,\alpha_a}_{k_1,\cdots,k_{a+2a'+1}}=\bar l_0 \prod_{i=1}^a
\tilde{\mathbb{W}}^{\alpha_i}_{k_i}\prod_{j=1}^{2a'+1}\tilde{\mathbb V}_{k_{a+j}},
\end{eqnarray}
in which
$\hat{\mathbb{W}}_{k_i}=k_i\frac{1-q^{k_i}}{1-t^{k_i}}\frac{\partial}{\partial g_{k_i}}-{\mathcal{\hat W}}_{-{k_i}}^{(n_1,n_2)}(\vec{u})$,
$\hat{\mathbb V}_{k_{a+j}}=\frac{1-q^{k_{a+j}}}{1-t^{k_{a+j}}}\frac{\partial}{\partial \tilde g_{k_{a+j}}}-{\mathcal{{\hat V}}}_{-{k_{a+j}}}^{(n_1,n_2)}(\vec{u})$, $\tilde{\mathbb{W}}_{k_i}={k_i}\frac{1-q^{k_i}}{1-t^{k_i}}\frac{\partial}{\partial h_{k_i}}-{\mathcal{{\hat W}}}_{{k_i}}^{(n_1,n_2)}(\vec{u})$, $\tilde{\mathbb V}_{k_{a+j}}=\frac{1-q^{k_{a+j}}}{1-t^{k_{a+j}}}\frac{\partial}{\partial \tilde h_{k_{a+j}}}-{\mathcal{{\hat V}}}_{{k_{a+j}}}^{(n_1,n_2)}(\vec{u})$,
$k_i,\alpha_i\in \mathbb{N}_+$ $(i=1,\cdots, a)$, $k_{a+j}\in \mathbb{N}$ $(j=1,\cdots, 2a'+1)$ and $a'\in \mathbb{N}$.
The constraint operators (\ref{dconopn}) and (\ref{dconopp}) satisfy the generalized super Virasoro algebra (\ref{gsvir})
and null super $3$-algebra (\ref{null3alg}).

We list the first several partition functions of (\ref{ehatzn}) as follows:
\begin{eqnarray}\label{exs}
%0,-1
\hat{\mathcal{Z}}_{0,-1}\{u;p,\tilde p|g, \tilde g\}&=&
\sum_{\lambda}
\frac{\mathcal{E}_{t^{u}}(P_{\lambda})}{\mathcal{E}'_{0}(P_{\lambda})}
P_{\lambda}\{p\}
Q_{\lambda}\{g\}+\sum_{\Lambda}\frac{\mathcal{\tilde E}_{t^{-m+1+u}}(P_{\Lambda})
\mathcal{E}_{t^{u}}(P_{\delta_{m}})}
{\mathcal{\tilde E}'_{0}(P_{\Lambda})\mathcal{E}'_{0}(P_{\delta_{m}})}\nonumber\\
&&\cdot P_{\Lambda}\{p,\tilde p\}Q_{\Lambda}\{g,\tilde g'\},\nonumber\\
%-1,0
\hat{\mathcal{Z}}_{-1,0}\{u;p,\tilde p|g, \tilde g\}
&=&\sum_{\lambda}
\frac{\mathcal{E}_{t^{u}}(P_{\lambda})}{\mathcal{E}'_{0}(P_{\lambda})}
P_{\lambda}\{p\}
Q_{\lambda}\{g\}+\sum_{\Lambda}
\frac{\mathcal{E}_{t^{-m+u}}(P_{\Lambda})\mathcal{E}_{t^{u}}(P_{\delta_{m+1}})}
{\mathcal{E}'_{0}(P_{\Lambda})\mathcal{E}'_{0}(P_{\delta_{m+1}})}
\nonumber\\
&&\cdot P_{\Lambda}\{p,\tilde p\}Q_{\Lambda}\{g,\tilde g'\},\nonumber\\
%-1,-1
\hat{\mathcal{Z}}_{-1,-1}\{\vec u;p,\tilde p|g, \tilde g\}&=&\sum_{\lambda}
\frac{\mathcal{E}_{t^{u_1}}(P_{\lambda})\mathcal{E}_{t^{u_2}}(P_{\lambda})}{(\mathcal{E}'_{0}(P_{\lambda}))^2}
P_{\lambda}\{p\}
Q_{\lambda}\{g\}+\sum_{\Lambda}
\frac{\mathcal{E}_{t^{-m+u_{1}}}(P_{\Lambda})}{\mathcal{E}'_{0}(P_{\Lambda})
}\nonumber\\
&&\cdot \frac{\mathcal{E}_{t^{u_{1}}}(P_{\delta_{m+1}})\mathcal{\tilde E}_{t^{-m+1+u_{2}}}(P_{\Lambda})
\mathcal{E}_{t^{u_{2}}}(P_{\delta_{m}})}
{\mathcal{E}'_{0}(P_{\delta_{m+1}})\mathcal{\tilde E}'_{0}(P_{\Lambda})\mathcal{E}'_{0}(P_{\delta_{m}})}
P_{\Lambda}\{p,\tilde p\}Q_{\Lambda}\{g,\tilde g'\},\nonumber\\
%0,1
\hat{\mathcal Z}_{0,1}\{u;p,\tilde p|g,\tilde g|h,\tilde h\}
&=&\sum_{\lambda,\mu}
\frac{\mathcal{E}_{t^{u}}(P_{\lambda})\mathcal{E}'_{0}(P_{\mu})}
{\mathcal{E}_{t^{u}}(P_{\mu})
\mathcal{E}'_{0}(P_{\lambda})}
P_{\lambda/\mu}\{h\}P_{\mu}\{p\}Q_{\lambda}\{g\}+\sum_{\Lambda,\mu}
\frac{\mathcal{\tilde E}_{t^{-m+1+u}}(P_{\Lambda})
\mathcal{E}_{t^{u}}(P_{\delta_m})}{\mathcal{\tilde E}'_{0}(P_{\Lambda})
\mathcal{E}'_{0}(P_{\delta_m})}\nonumber\\
&&\cdot \frac{\mathcal{E}'_{0}(P_{\mu})}{\mathcal{E}_{t^{u}}(P_{\mu})}P_{\Lambda/\mu}\{h,\tilde h'\}P_{\mu}\{p\}Q_{\Lambda}\{g,\tilde g'\}
+\sum_{\Lambda,\Gamma}
\frac{\mathcal{\tilde E}_{t^{-m+1+u}}(P_{\Lambda})\mathcal{\tilde E}'_{0}(P_{\Gamma})}
{\mathcal{\tilde E}_{t^{-m'+1+u}}(P_{\Gamma})\mathcal{\tilde E}'_{0}(P_{\Lambda})}
\nonumber\\
&&\cdot \frac{\mathcal{E}_{t^{u}}(P_{\delta_{m}})
\mathcal{E}'_{0}(P_{\delta_{m'}})}{\mathcal{E}_{t^{u}}(P_{\delta_{m'}})
\mathcal{E}'_{0}(P_{\delta_{m}})}P_{\Lambda/\Gamma}\{h,\tilde h'\}P_{\Gamma}\{p,\tilde p\}Q_{\Lambda}\{g,\tilde g'\},\nonumber\\
%1,0
\hat{\mathcal Z}_{1,0}\{u;p,\tilde p|g,\tilde g|h,\tilde h\}
&=&\sum_{\lambda,\mu}
\frac{\mathcal{E}_{t^{u}}(P_{\lambda})\mathcal{E}'_{0}(P_{\mu})}
{\mathcal{E}_{t^{u}}(P_{\mu})
\mathcal{E}'_{0}(P_{\lambda})}
P_{\lambda/\mu}\{h\}P_{\mu}\{p\}Q_{\lambda}\{g\}+\sum_{\Lambda,\mu}
\frac{\mathcal{E}_{t^{-m+u}}(P_{\Lambda})
\mathcal{E}_{t^{u}}(P_{\delta_{m+1}})}
{\mathcal{E}'_{0}(P_{\Lambda})
\mathcal{E}'_{0}(P_{\delta_{m+1}})}\nonumber\\
&&\cdot \frac{\mathcal{E}'_{0}(P_{\mu})}{\mathcal{E}_{t^{u}}(P_{\mu})}P_{\Lambda/\mu}\{h,\tilde h'\}P_{\mu}\{p\}Q_{\Lambda}\{g,\tilde g'\}
+\sum_{\Lambda,\Gamma}
\frac{\mathcal{E}_{t^{-m+u}}
(P_{\Lambda})\mathcal{E}'_{0}(P_{\Gamma})}{\mathcal{E}_{t^{-m'+u}}
(P_{\Gamma})\mathcal{E}'_{0}
(P_{\Lambda})}\nonumber\\
&&\cdot\frac{\mathcal{E}_{t^{u}}(P_{\delta_{m+1}})
\mathcal{E}'_{0}(P_{\delta_{m'+1}})}
{\mathcal{E}_{t^{u}}(P_{\delta_{m'+1}})
\mathcal{E}'_{0}(P_{\delta_{m+1}})}
P_{\Lambda/\Gamma}\{h,\tilde h'\}P_{\Gamma}\{p,\tilde p\}Q_{\Lambda}\{g,\tilde g'\},\nonumber\\
%1,1
\hat{\mathcal Z}_{1,1}\{\vec u;p,\tilde p|g,\tilde g|h,\tilde h\}
&=&\sum_{\lambda,\mu}
\frac{\mathcal{E}_{t^{u_1}}(P_{\lambda})\mathcal{E}_{t^{u_2}}(P_{\lambda})(\mathcal{E}'_{0}(P_{\mu}))^2}
{\mathcal{E}_{t^{u_1}}(P_{\mu})\mathcal{E}_{t^{u_2}}(P_{\mu})
(\mathcal{E}'_{0}(P_{\lambda}))^2}
P_{\lambda/\mu}\{h\}P_{\mu}\{p\}Q_{\lambda}\{g\}
\nonumber\\
&&+\sum_{\Lambda,\mu}
(\mathcal{E}'_{0}(P_{\mu}))^{2}
\frac{\mathcal{E}_{t^{-m+u_{1}}}(P_{\Lambda})
\mathcal{E}_{t^{u_{1}}}(P_{\delta_{m+1}})
\mathcal{\tilde E}_{t^{-m+1+u_{2}}}(P_{\Lambda})
\mathcal{E}_{t^{u_{2}}}(P_{\delta_m})}
{\mathcal{E}'_{0}(P_{\Lambda})\mathcal{E}_{t^{u_{1}}}(P_{\mu})
\mathcal{E}'_{0}(P_{\delta_{m+1}})\mathcal{\tilde E}'_{0}(P_{\Lambda})\mathcal{E}_{t^{u_{2}}}(P_{\mu})
\mathcal{E}'_{0}(P_{\delta_m})}\nonumber\\
&&\cdot P_{\Lambda/\mu}\{h,\tilde h'\}P_{\mu}\{p\}Q_{\Lambda}\{g,\tilde g'\}
+\sum_{\Lambda,\Gamma}
\frac{\mathcal{E}_{t^{-m+u_{1}}}
(P_{\Lambda})\mathcal{E}'_{0}(P_{\Gamma})}{\mathcal{E}_{t^{-m'+u_{1}}}
(P_{\Gamma})\mathcal{E}'_{0}
(P_{\Lambda})}
\nonumber\\
&&\cdot \frac{\mathcal{E}_{t^{u_{1}}}(P_{\delta_{m+1}})
\mathcal{E}'_{0}(P_{\delta_{m'+1}})\mathcal{\tilde E}_{t^{-m+1+u_{2}}}(P_{\Lambda})\mathcal{\tilde E}'_{0}(P_{\Gamma})\mathcal{E}_{t^{u_{2}}}(P_{\delta_{m}})
\mathcal{E}'_{0}(P_{\delta_{m'}})}
{\mathcal{E}_{t^{u_{1}}}(P_{\delta_{m'+1}})
\mathcal{E}'_{0}(P_{\delta_{m+1}})\mathcal{\tilde E}_{t^{-m'+1+u_{2}}}(P_{\Gamma})\mathcal{\tilde E}'_{0}(P_{\Lambda})\mathcal{E}_{t^{u_{2}}}(P_{\delta_{m'}})
\mathcal{E}'_{0}(P_{\delta_{m}})}
\nonumber\\
&&\cdot P_{\Lambda/\Gamma}\{h,\tilde h'\}P_{\Gamma}\{p,\tilde p\}Q_{\Lambda}\{g,\tilde g'\}.
\end{eqnarray}

It is clear that the $\beta$-deformed cases in the previous section can be recovered by taking the limit $t=q^{\beta}$, $q\rightarrow 1$ in (\ref{ehatzn}).
Furthermore, setting to zero all the fermionic variables in (\ref{ehatzn}), we arrive at the results in Ref. \cite{qtHurwitz}
\begin{eqnarray}\label{zmu}
  {\hat Z}_{-n}\{\vec{u};p,g\}
  &=&\sum_{\lambda} \prod_{i=1}^{n}\frac{P_{\lambda} \{p_k=\pi_{k}^{(u_i)}\}}
  {P_{\lambda}\{p_k=\delta_{k,1}^{*}\}}
  P_{\lambda}\lbrace p\rbrace Q_{\lambda}\lbrace g\rbrace,\label{nzn}\\
{\hat{Z}}_{n}\{\vec{u};\bar{p},p,g\}&=&\sum_{\lambda,\mu} \prod_{i=1}^{n}\frac{P_{\lambda} \{p_k=\pi_{k}^{(u_i)}\}	 P_{\mu}\{p_k=\delta_{k,1}^{*}\}}
	{P_{\lambda}\{p_k=\delta_{k,1}^{*}\} P_{\mu}\{p_k=\pi_{k}^{(u_i)}\}}P_{\lambda /\mu}\{\bar{p}\} P_{\mu}\{p\} Q_{\lambda}\{g\},\label{znm}
 \end{eqnarray}
where $n=n_1+n_2$, $\pi_{k}^{(u_i)}=\frac{1-t^{ku_i}}{1-t^{k}}$ and
$\delta_{k,1}^{*}=\frac{(1-q)^{k}}{1-t^{k}}$.
It is noted that for the $(q,t)$-deformed partition functions (\ref{nzn}) and (\ref{znm}) with given $n$, there are
 $n+1$ supersymmetric extensions, respectively.

The $(q,t)$-deformed Gaussian hermitian model is given by \cite{Mironovsummary,Morozov2018,Morozov}
\begin{eqnarray}
   {\hat{Z}}_{-1}\{N;p_k=\delta_{k,2}^{*},g\}
   &=&\prod_{i=1}^{N} \int_{-1}^1{d_q}x_i x_{i}^{\beta(N-1)} (q(1-q)^{\frac{1}{2}}x_i;q)_{\infty}(-q(1-q)^{\frac{1}{2}}x_i;q)_{\infty} \nonumber\\
&&\cdot \prod_{1\leq k\neq l \leq N}\frac{(x_{k}/x_{l};q)_{\infty}}{(tx_{k}/x_{l};q)_{\infty}}
e^{\sum_{s \geq 1}\frac{1-t^s}{1-q^s}\frac{p_s}{s}\sum_{i=1}^{N}x_i^s}\nonumber\\
 &=&\sum_{\lambda} \frac{P_{\lambda} \{p_k=\pi_{k}^{(N)} \}} {P_{\lambda}\{p_k=\delta_{k,1}^{*}\}}
   P_{\lambda}\{p_k=\delta_{k,2}^{*}\} Q_{\lambda}\{g\}, \label{ztn1}
\end{eqnarray}
where the integral is given by the Jackson integral, $\delta_{k,2}^{*}=\frac{2(1-q)^{k\over 2}}{1-t^k}\delta_{k|2}$.
We see that $\hat{\mathcal{Z}}_{0,-1}\{N;p_k=\delta_{k,2}^{*},\tilde p|g, \tilde g\}$ and $\hat{\mathcal{Z}}_{-1,0}\{N;p_k=\delta_{k,2}^{*},\tilde p|g, \tilde g\}$
are the supersymmetric extensions of character expansion of (\ref{ztn1}).

For the supersymmetric partition functions $\hat{\mathcal{Z}}_{0,-2}\{N_1,N_2;p_k=\delta_{k,1}^{*}, \tilde p|g, \tilde g\}$,
$\hat{\mathcal{Z}}_{-2,0}\{N_1,N_2;p_k=\delta_{k,1}^{*},\tilde p|g, \tilde g\}$
and $\hat{\mathcal{Z}}_{-1,-1}\{N_1,N_2;p_k=\delta_{k,1}^{*},\tilde p|g, \tilde g\}$, they
may be thought of as the supersymmetric extensions of $(q,t)$-deformed complex model \cite{Morozov2018,Morozov}
\begin{eqnarray}
{\hat{Z}}_{-2}\{N_1,N_2;p_k=\delta_{k,1}^{*},g\}=
\sum_{\lambda} \frac{P_{\lambda}\{p_k=\pi_{k}^{(N_1)}\}P_{\lambda}\{p_k=\pi_{k}^{(N_2)}\}} {P_{\lambda}\{p_k=\delta_{k,1}^{*}\}}Q_{\lambda}\{g\}.
\end{eqnarray}

The partition function for the $3d$ $\mathcal{N}=2$ supersymmetric gauge theories on $D^2\times_q S^1$ with $N_f=1$ is given by \cite{Cassia}
\begin{eqnarray}\label{4dpf}
&&{\hat{Z}}_{-2}\{{\rm log}_t^q +N-1,N;p_k=(1-q)^k\delta_{k,1}^{*},g\}\nonumber\\
&=&\prod_{i=1}^{N} \int_{0}^1{d_q}x_i x_{i}^{\beta(N-1)} (qx_i;q)_{\infty} \prod_{1\leq k\neq l \leq N}\frac{(x_{k}/x_{l};q)_{\infty}}
{(tx_{k}/x_{l};q)_{\infty}}e^{ \sum_{s \geq 1}\frac{1-t^s}{1-q^s}\frac{p_s}{s}\sum_{i=1}^{N}x_i^s},
\end{eqnarray}
where
$\beta=\log_{q}{t}$, $(x;q)_{\infty}=\prod_{i=1}^{\infty}(1-q^{i-1}x)$.

A remarkable property of (\ref{4dpf})
is the superintegrablity. More precisely, it can be expressed as the following character expansion
\begin{eqnarray}\label{4dpfce}
&&{\hat{Z}}_{-2}\{{\rm log}_t^q +N-1,N;p_k=(1-q)^k\delta_{k,1}^{*},g\}\nonumber\\
&=&\sum_{\lambda} \frac{P_{\lambda}\{p_k=\frac{1-q^kt^{(N-1)k}}{1-t^{k}}\}P_{\lambda}\{p_k=\pi_{k}^{(N)}\}} {P_{\lambda}\{p_k=\frac{1}{1-t^{k}}\}}Q_{\lambda}\{g\}.
\end{eqnarray}
We see that the supersymmetric extensions of (\ref{4dpfce}) are
$\hat{\mathcal{Z}}_{0,-2}\{{\rm log}_t^q +N-1,N;p_k=(1-q)^k\delta_{k,1}^{*}, \tilde p|g, \tilde g\}$,
$\hat{\mathcal{Z}}_{-2,0}\{{\rm log}_t^q +N-1,N;p_k=(1-q)^k\delta_{k,1}^{*},\tilde p|g, \tilde g\}$
and $\hat{\mathcal{Z}}_{-1,-1}\{{\rm log}_t^q +N-1,N;p_k=(1-q)^k\delta_{k,1}^{*},\tilde p|g, \tilde g\}$.

\section{Conclusions}
%%%%%%%%%%%%%%%%%%%%%%%%%%%%%%%%%%%%%%%%%%%%%%%%%%%%%%%%%%%
We have extended the $\beta$ and $(q,t)$-deformed Hurwitz operators to the supersymmetric cases. Then we gave
the supersymmetric $\beta$ and $(q,t)$-deformed HK partition functions (\ref{SHKM}) and (\ref{2qtSHKM}) through $W$-representations.
The superintegrability was shown by the character expansions with respect to the Jack and Macdonald superpolynomials,
respectively.

Based on the $\beta$-deformed Hurwitz superoperators (\ref{shkope}), we have constructed a series of bosonic and fermionic operators
by the nested commutators. Then we gave the supersymmetric $\beta$-deformed hierarchy (see negative and positive branches
(\ref{eznm}) and (\ref{eszngp})) for the partition functions through $W$-representations,
where the $W$-operators in $W$-representations were given by the constructed bosonic and fermionic operators. By providing the corresponding
character expansions with respect to the Jack superpolynomials, the superintegrablity for supersymmetric $\beta$-deformed partition function
hierarchy has been confirmed, i.e., $<character>\sim character$. Moreover, when $\beta=1$, the Jack superpolynomials in the character expansions
reduce to the Schur-Jack superpolynomials. The superintegrablity for the reduced supersymmetric hierarchy still holds.
In addition, we have constructed the generalized super Virasoro constraints (\ref{vircons1}) and (\ref{vircons2}) for the supersymmetric $\beta$-deformed
partition functions, where the constraint operators obey the generalized super Virasoro algebra and null super 3-algebra. From the algebraic point of view,
the new infinite-dimensional super (3-)algebras presented in this paper are also interesting in their own right due to the super higher algebraic structures.
Searching for other realizations of such super (3-)algebras and more applications would be interesting.

Similarly, based on the $(q,t)$-deformed superoperators (\ref{O1}), we have also constructed the supersymmetric $(q,t)$-deformed partition function
hierarchy (\ref{ehatzn}) through $W$-representations and presented the generalized super Virasoro constraints (\ref{vircons3}) and (\ref{vircons4}).
The superintegrability was shown by the character expansions with respect to the Macdonald superpolynomials. When we set all the fermionic variables to zero
in the supersymmetric ($\beta$ and $(q,t)$-deformed) partition functions presented in this paper, they coincide with the known results
in the literature \cite{2301.11877,2301.12763,qtHurwitz}. Our results may shed new light on supereigenvalue and supermatrix models. For further research,
it would be interesting to search for the integral representations for the supersymmetric ($\beta$ and $(q,t)$-deformed) partition functions with $W$-representations.

\section *{Acknowledgments}
We are grateful to A. Morozov, A. Mironov, V. Mishnyakov and A. Popolitov for their helpful comments.
This work is supported by the National Natural Science Foundation of China (Nos. 12205368 and 11875194)
and the Fundamental Research Funds for the Central Universities, China (No. 2022XJLX01).

%%%%%%%%%%%%%%%%%%%%%%%%%%%%%%%%%%%%%%%%%%%%%%%%%%%%%%%%%%%%%%%%%%%%%%%%%%%%%%%

\end{document}